\documentclass[aps,prb,twocolumn,showpacs,floatfix,superscriptaddress]{revtex4-1}
\usepackage{dcolumn}
\usepackage{bm}
\usepackage{amsmath,amssymb,graphicx}
\usepackage[colorlinks=true,citecolor=blue,urlcolor=blue,linkcolor=blue]{hyperref}

\begin{document}
\preprint{Submitted to Phys. Rev. B}

\title{Terahertz radiation from accelerating charge carriers in graphene under ultrafast photoexcitation}

\author{Avinash Rustagi}
\author{C.~J.~Stanton}
\email[]{avinash@phys.ufl.edu}
\thanks{corresponding author.}
\affiliation{Department of Physics, University of Florida, Gainesville, Florida, 32611}

\date{\today}
\begin{abstract}
We study the generation of THz radiation from the acceleration of ultrafast photoexcited charge carriers in graphene in the presence of a DC electric field. Our model is based on calculating the transient current density from the time-dependent distribution function which is determined using the Boltzmann transport equation within a relaxation time approximation. We include the time-dependent generation of carriers by the pump pulse by solving for the carrier generation rate using the Bloch equations in the rotating wave approximation (RWA). The linearly polarized pump pulse generates an anisotropic distribution of photoexcited carriers in the $k_x-k_y$ plane. The collision integral in the Boltzmann equation includes a term that leads to the \textit{thermalization} of carriers via carrier-carrier scattering to an effective temperature above the lattice temperature, as well as a \textit{cooling} term which leads to energy relaxation via inelastic carrier-phonon scattering. The radiated signal is proportional to the time derivative of the transient current density. In spite of the fact that the magnitude of the velocity is the same for all the carriers in graphene, there is still emitted radiation from the photoexcited charge carriers with frequency components in the THz range due to a change in the \textit{direction} of velocity of the photoexcited carriers in the external electric field as well as \textit{cooling} of the photoexcited carriers on a sub-picosecond time scale.
\end{abstract}

\pacs{78.67.-n,78.20.Bh,73.50.Gr}
\maketitle

\section{Introduction}

Terahertz (THz) radiation lies in the 0.1-10 THz frequency range and this frequency range is referred to as the Terahertz gap. The frequency region below 0.1 THz is where conventional electronics operates and the region above 10 THz belongs to optics. This has been one of the reasons for a dearth of generating sources and detectors for THz radiation since it cannot be detected or generated by conventional electronics used in the case of radio waves and microwaves and novel methods continue to make advances in this field. Imaging, sensing and spectroscopy using THz radiation has widespread applications in a variety of contexts\cite{Maagt_THzBook}. Imaging and sensing using THz radiation has promising applications in tumor detection for biomedical applications, inspection/process control in industry, studying cosmic background radiation in astronomy, explosive detection for security purposes, remote sensing and ultrafast wireless communication. Spectroscopy using THz radiation is ideally suited to study low energy excitations like superconducting gap in superconductors, ionization energies of shallow donors and acceptors in semiconductors, spin flip energies, etc.\\

Photonics based devices are used for generation of high THz frequencies close to optical frequencies, like the free electron laser (FEL)\cite{Tan2012FEL} and quantum cascade laser (QCL)\cite{THzQCL}. FEL THz source includes an electron accelerator and an undulator which produces a magnetic field that accelerates the electron beam and radiates in the THz range. On the other hand, QCL has a cascade of quantum wells which have sub-bands due to confinement effects where the energy spacing between the sub-bands determine the lasing frequency. Electronics based devices are also used for generation of low THz frequencies close to electronic device frequencies, like Gunn oscillators\cite{alekseev2000gan} and optical parametric oscillators\cite{pala2012terahertz}. Gunn diode oscillators are based on the negative differential resistance regime at high electric field strengths where intervalley electron transfer becomes significant. Optical parametric oscillators on the other hand are based on the optical gain from parametric amplification of a non-linear crystal.  \\

Semiconductors have been used to generate THz radiation through several mechanisms wherein ultrashort pulses are incident on a semiconductor surface and a THz pulse reflects off alongside a portion of the incident pulse. The main mechanisms for generation of THz radiation from semiconductors are:

\begin{enumerate}
\item Optical non-linearity of the material causing radiation through higher order susceptibility $\chi_2$\cite{Tonouchi2007,MaAndZhang1993,RiceMaAndZhang1994}.
\item Virtual carriers by photo-exciting carrier below the band gap in presence of DC electric field\cite{KuznetsovStanton1993,KuznetsovStantonAuston1994}.
\item Photo-Dember effect where a dipole moment forms close to the surface due to the difference in mobilities/diffusion of electrons and holes\cite{Dekorsy1996,Johnston2002}.
\item Photoexcited charge carrier acceleration in presence of an internal or external electric field\cite{ZhangAuston1992,Dekorsy1993}.
\end{enumerate}

Graphene is a zero-gap semiconductor with vanishing density of states at the Dirac points\cite{novoselov2005two}. The linearity of the energy dispersion about the Dirac point has interesting consequences in both transport and optical measurements namely the presence of a zero energy Landau level\cite{Booshehri:2012CR}, anomalous quantum hall effect\cite{Guysnin2005_QHE,Zhang2005_QHE}, good hot electron noise property due to velocity fluctuations\cite{Rustagi2014} among others{\cite{CastroNetoRMP}}. This linearity of the dispersion near the Dirac points along with the high carrier velocity causes large nonlinear optical response\cite{GlazovGanichev2014}. Second order nonlinearity from anisotropic photoexcitation by \textit{oblique} optical pulse excitation can generate THz frequencies in graphene\cite{Mangeney2014}.

\begin{figure}[htbp]
\centering
\includegraphics[width=0.45\textwidth]{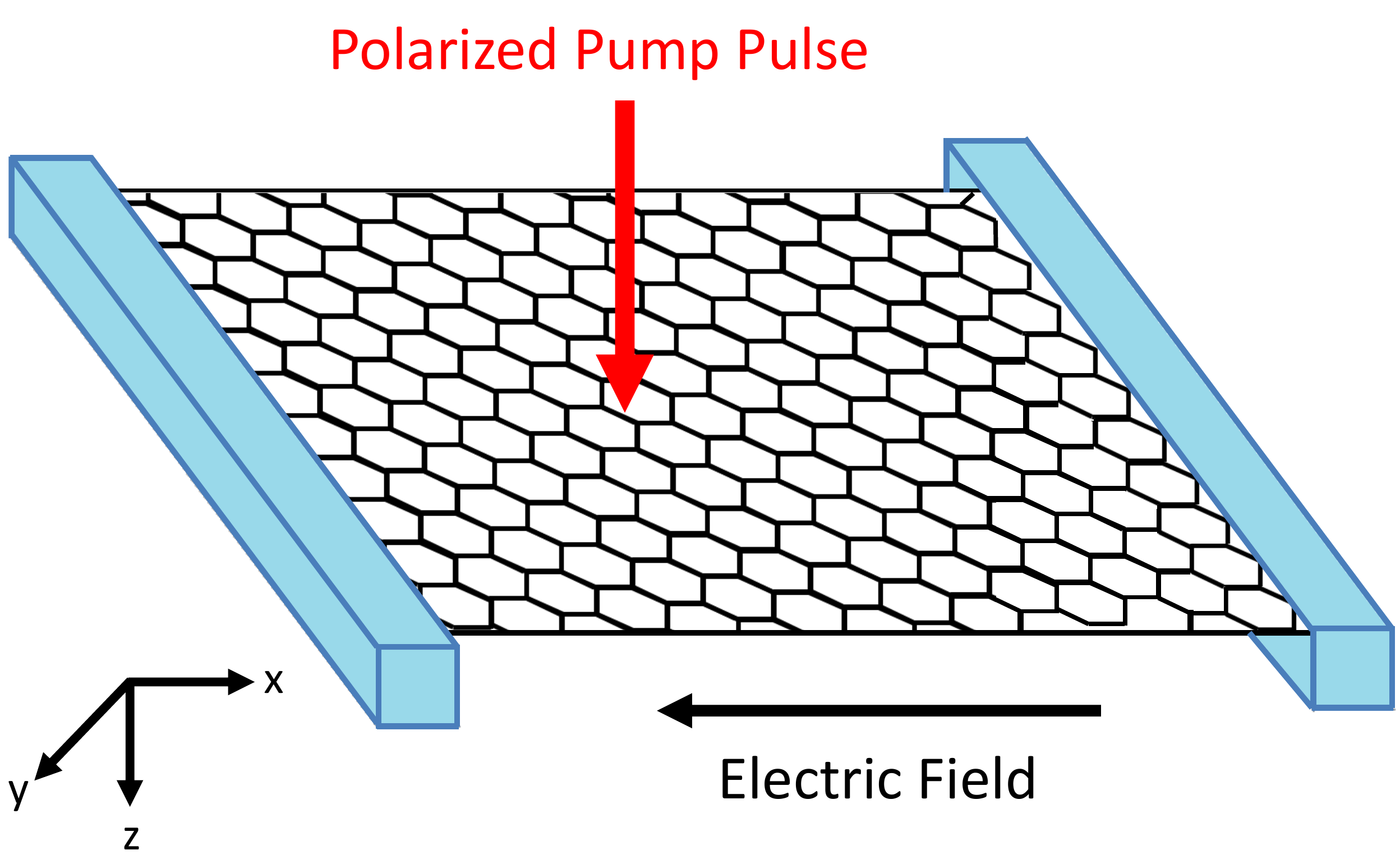}
\caption{\label{configuration}(color online) Suspended graphene with a DC electric field applied in the graphene plane and polarized pump pulse incident perpendicular to the sheet.}
\end{figure}

In this paper, we are interested in investigating THz generation from ultrafast photoexcited carriers in graphene, with emphasis on charge carrier acceleration. We take the external electric field to be in-plane rather than perpendicular as is usually the case (Fig.~{\ref{configuration}}). Because the graphene crystal structure is centro-symmetric, we expect no optical non-linearity contribution without excitation at \textit{oblique} incidence. Also, since the valence and conduction bands in graphene are symmetric about the Dirac point for low energies (implying similar electron and hole mobilities) in graphene, the Photo-Dember effect should be small\cite{Graphene_Dember}. One might also expect that THz generation through charge carrier acceleration would also be small, since the magnitude of the velocity is \textit{constant} in graphene. However, the \textit{direction} of the velocity changes which leads to THz radiation.

In Sec.~\ref{Simplistic Model}, we explain the key difference in THz generation process between the conventional parabolic dispersion and the Dirac dispersion in a simple model. Then we present a more detailed model used in this manuscript in Sec.~\ref{Model}. In Sec.~{\ref{BTE}}, we formulate the model using the time-dependent Boltzmann equation in the relaxation time approximation. In Sec.~{\ref{Results}}, we summarize the results for the undoped graphene and n-doped graphene with room temperature chemical potential of 50 meV. 

\begin{figure}[htbp]
\centering
\includegraphics[width=0.49\textwidth]{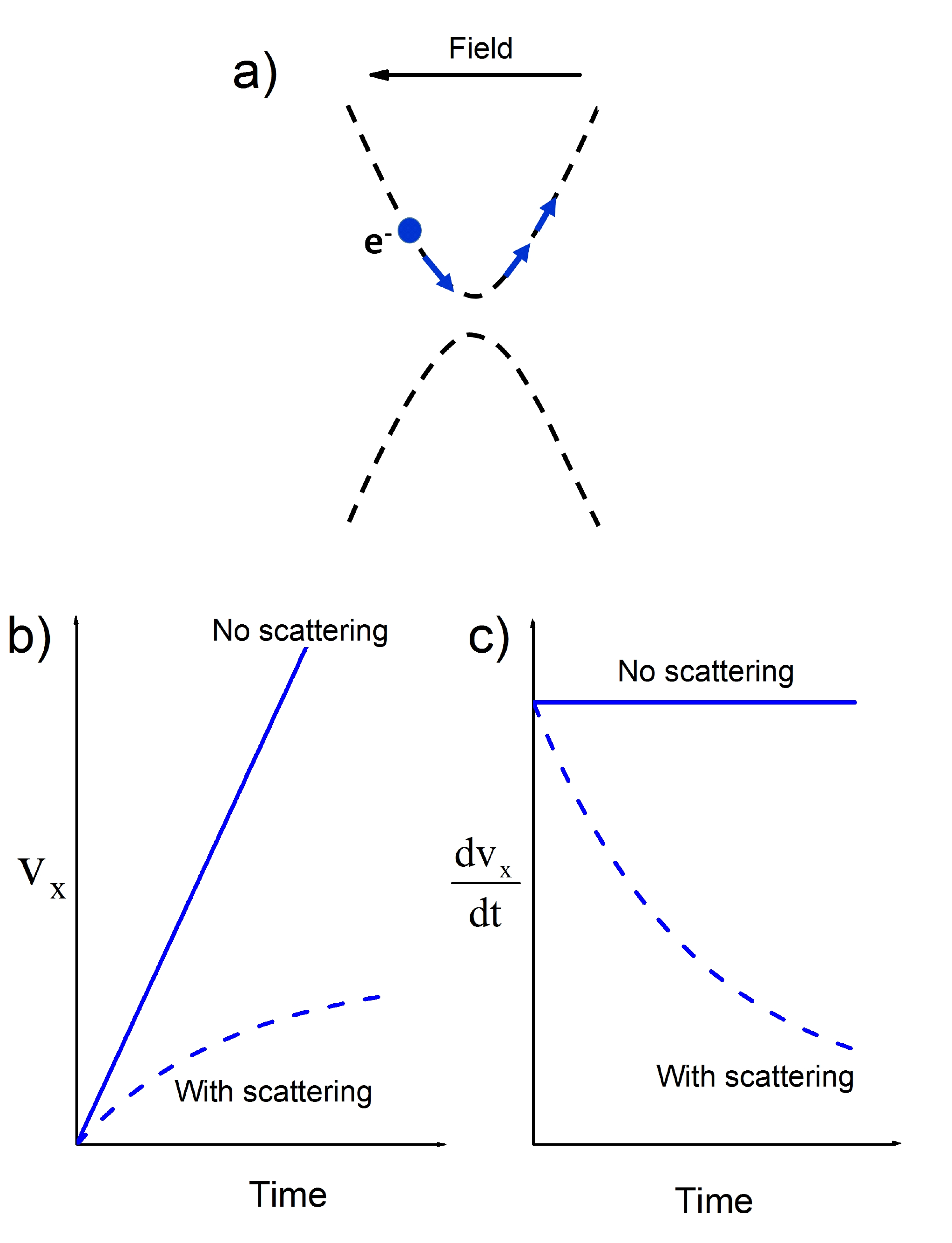}
\caption{\label{Parabolic}(color online) a) Schematic of time evolution of wave-vector of an electron in parabolic dispersion, b) Component of velocity along the field direction of a carrier and c) time derivative of velocity as a function of time in presence and absence of scattering.}
\end{figure}

\section{Simplistic Model}
\label{Simplistic Model}
For a typical semiconductor, the velocity of a carrier is proportional to the wavevector $\bm{v} \propto \bm{k}$. This implies that the current due to this carrier in an electric field continues to increase linearly with time in absence of momentum relaxation mechanisms. The time derivative of the velocity gives the acceleration which is a constant in absence of scattering, implying constant radiation. Momentum scattering mechanisms  are important here since it limits the increase in velocity and there is a time varying acceleration as shown in Fig.~\ref{Parabolic}.

\begin{figure}[htbp]
\centering
\includegraphics[width=0.49\textwidth]{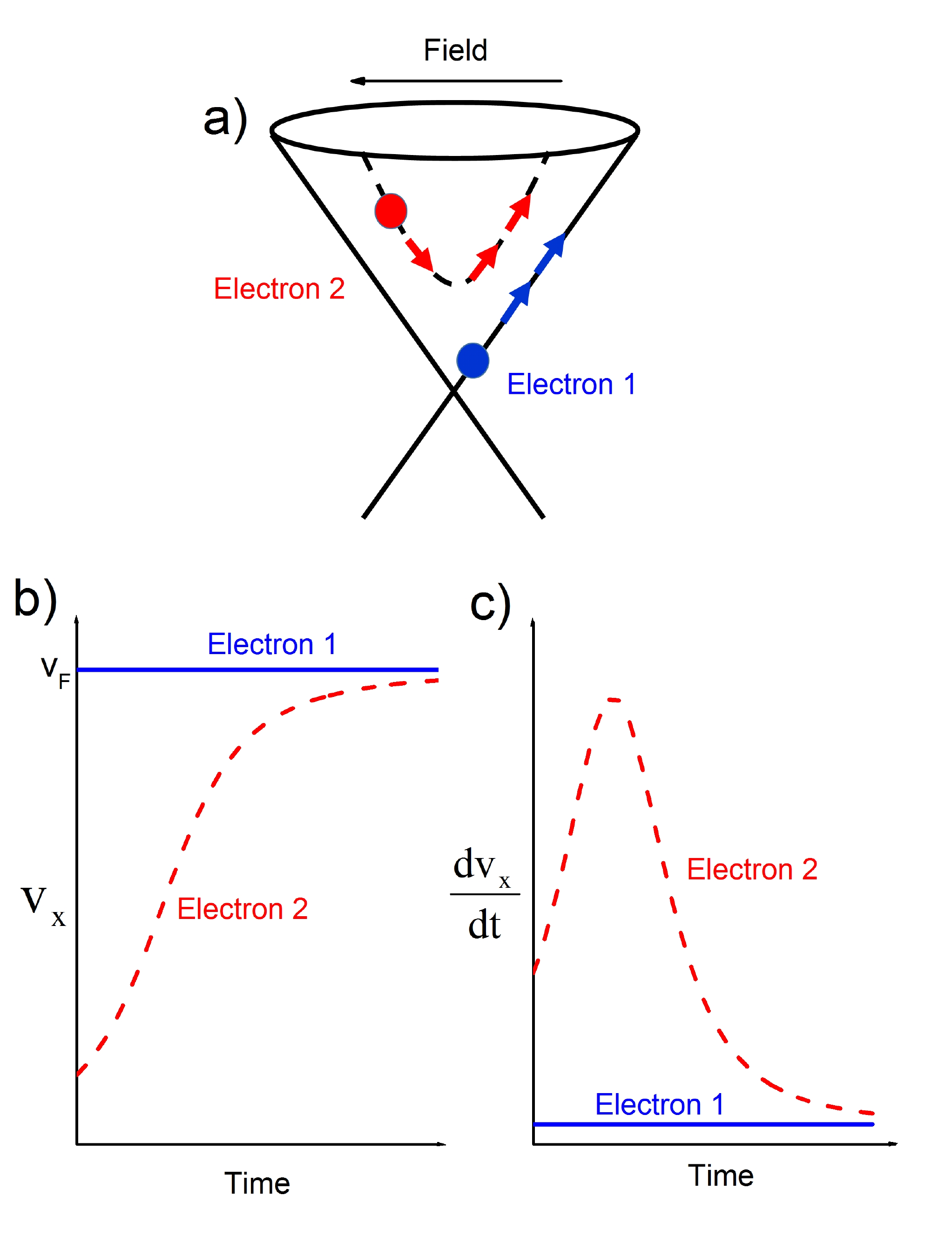}
\caption{\label{Graphene_Toy}(color online) a) Schematic of time evolution of wave-vector of an electron in graphene Dirac dispersion, b) Component of velocity along the field direction of a carrier and c) time derivative of velocity as a function of time in absence of scattering for two different initial wave-vectors.}
\end{figure}

The linear energy dispersion in graphene is much different from the otherwise conventional quadratic dispersion ($\varepsilon \propto k^2$). At the outset, it appears that the velocity is a constant for linear dispersion. However, the eigenstate of the electron matters when it comes to acceleration, for example there is no radiation for `Electron 1' as shown in Fig.~\ref{Graphene_Toy}. The direction of the velocity vector changes for `Electron 2' and this change in direction implies a time dependent acceleration as shown  in Fig.~\ref{Graphene_Toy}. Thus graphene has an interesting time dependent acceleration even in absence of scattering. The effects of scattering make a more detailed model and will be discussed in later section.\\

An electron in graphene in presence of external electric field has time varying $x$ and $y$ components of velocity. From the  semiclassical equation of motion, the time varying velocity can be calculated
\begin{equation}
\bm{v}=v_{F} \dfrac{\bm{k}-e\bm{E}t/\hbar}{\vert \bm{k}-e\bm{E}t/\hbar  \vert}.
\end{equation}
which indicates that the variation of the velocity of an electron depends on its initial $\bm{k}$-state. $e>0$ is the magnitude of electronic charge, $\bm{E}$ is the electric field, $v_F$ is the Fermi velocity of graphene, $t$ is the time.
%
%
\section{Model}
\label{Model}

The configuration shown in Fig.~{\ref{configuration}} displays a graphene sheet in the $x-y$ plane in presence of a DC electric field applied along the plane. A pump pulse is applied perpendicular to the graphene sheet which photo-excites carriers in the conduction band and holes in the valence band. These carriers accelerate in presence of the DC field giving rise to a time-dependent photocurrent. The time derivative of this photocurrent is proportional to the radiated signal. \\

\begin{figure}[htbp]
\centering
\includegraphics[width=0.495\textwidth]{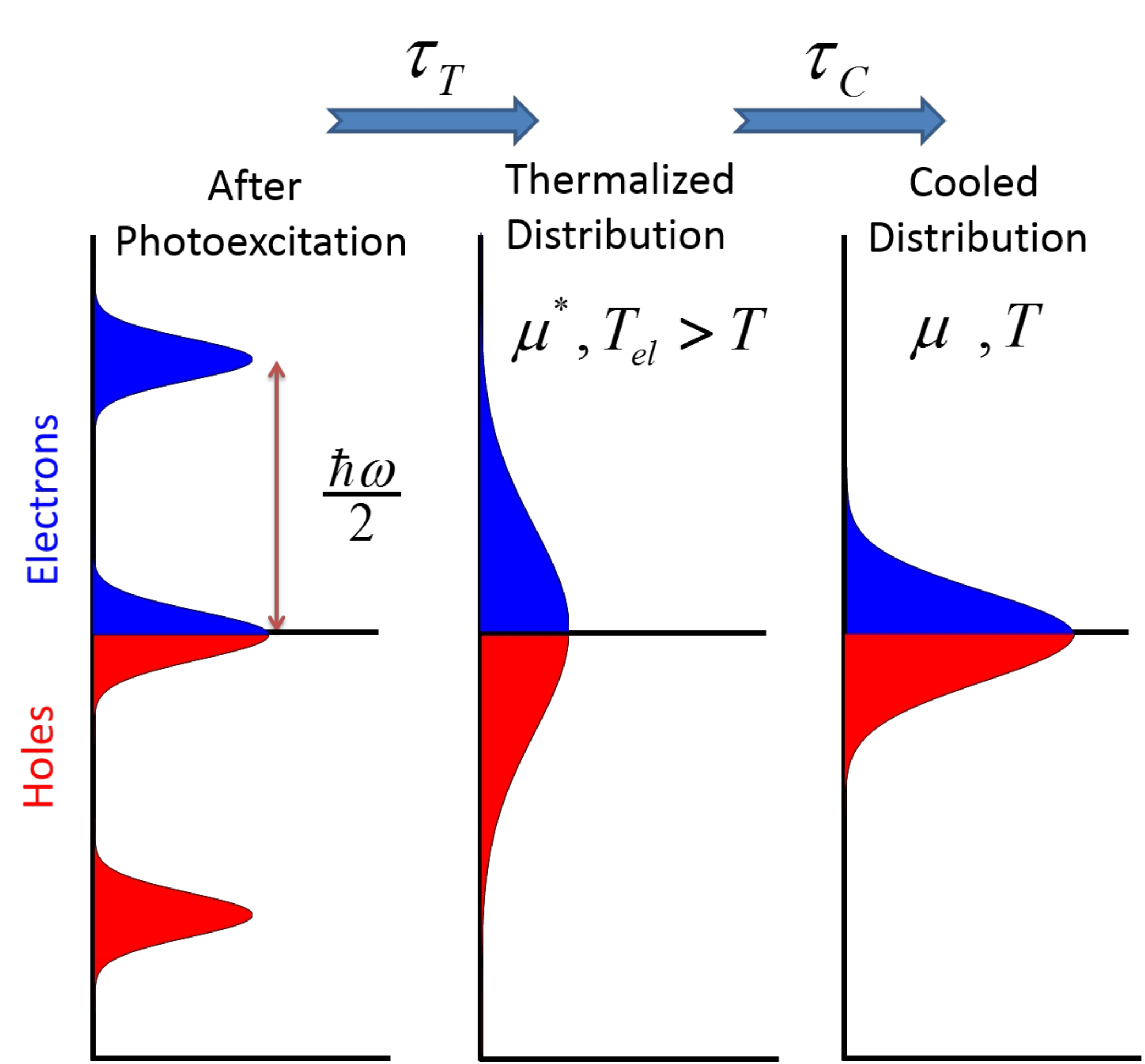}
\caption{\label{relaxation}(color online) Relaxation mechanisms include a `Thermalization' term and a `Cooling' term. The initial carriers along with the photo-excited carriers thermalize to an effective drifted `Thermalized Fermi-Dirac' distribution over a time scale $\tau_T$ and eventually cool to a `Cooled Fermi-Dirac' distribution over a time scale $\tau_C$ .}
\end{figure}

The most important relaxation mechanisms include a `Thermalization' term and a `Cooling' term (Fig.~\ref{relaxation}). The initial carriers along with the photo-excited carriers thermalize to an effective drifted `Thermalized Fermi-Dirac' distribution over a time scale $\tau_T$ with a drift wavevector $k_d$, a quasi-chemical potential $\mu^*$ and an effective carrier temperature $T_{el}$ which is higher than the lattice temperature $T$ because of energy being pumped into the system by photo-excitation. This eventually cools to a `Cooled Fermi-Dirac' distribution over a time scale $\tau_C$ with a chemical potential $\mu$ at the lattice temperature $T$. The recombination of carriers across the bands is ignored since it occurs on a much longer time scale compared to the thermalization and cooling times.

To model all of this, the time-dependent Boltzmann equation of carriers in presence of a  DC electric field is solved within the constant relaxation time approximation including a `Thermalization' term and a `Cooling' term. A carrier generation term is also included in the Boltzmann equation to take into account the time-dependent generation of carriers by the pump pulse. To get analytic results, the pump excitation is assumed to be weak and an expression for the generation rate of carriers is obtained by solving the Optical Bloch equations in the Rotating Wave approximation (RWA). Note that just electrons are considered in this model since holes will have the same contribution to the final current density.


\subsection{Boltzmann equation formulation}
\label{BTE}

The generation rate of the carriers will be used to study the transient current density from the Boltzmann Transport Equation(BTE) in presence of an in-plane DC electric field ($\vec{E}=-E\hat{x}$, $\vec{F}=-e\vec{E}=eE\hat{x}$).
\begin{equation}
\label{BTEGeneration}
\begin{split}
\dfrac{\partial f}{\partial t}+\dfrac{eE}{\hbar}\dfrac{\partial f}{\partial k_x}&=\dfrac{\partial f_g}{\partial t}\bigg\vert_{Generation}+I_{collision}\{f\} \\
&=\dfrac{\partial f_g}{\partial t}\bigg\vert_{Generation}-\dfrac{f-f_t}{\tau_t}-\dfrac{f-f_c}{\tau_c}
\end{split}
\end{equation}

We neglect the recombination of the carriers across the band since the typical time scale corresponding to this mechanism is much larger than the $ps$-time scale.\\

\begin{figure*}
\centering
\includegraphics[width=6 in,height=3.5 in]{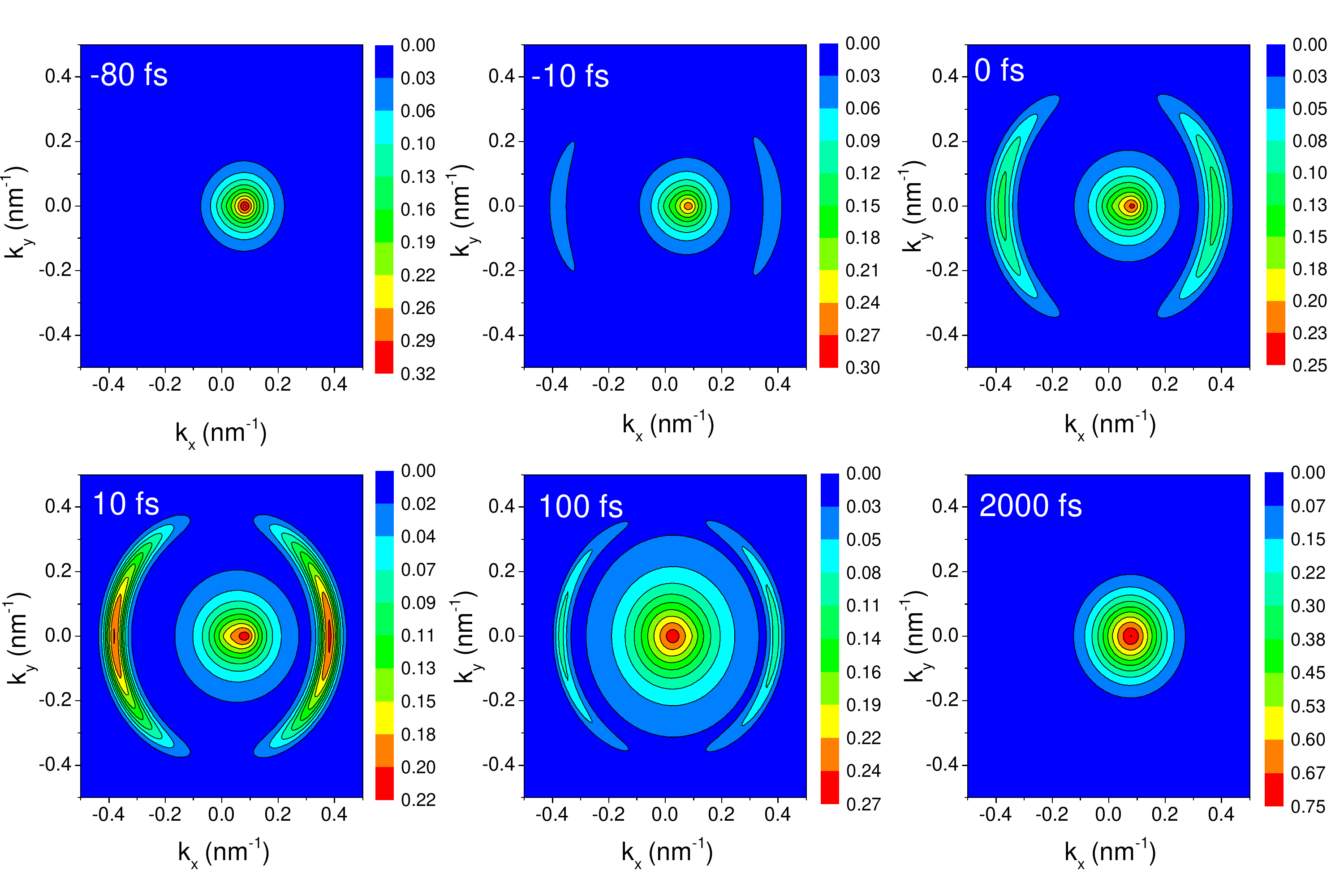}
\caption{\label{contour_0meV} (color online) Time evolution of the distribution function for undoped graphene. There is a non-zero, thermally excited carrier density (at room temperature) in the conduction band. The carriers are generated by the pump pulse with $Y$-polarization ( $\propto \cos^2 \theta$) during the pulse width duration. The carriers then thermalize to a hot Fermi-Dirac distribution which eventually cools down to the lattice temperature.}
\end{figure*}

The generation rate term corresponds to the carrier generation due to the ultrafast pump pulse evaluated using the Bloch equations in the weak pump limit as described in Appendix~\ref{Bloch Equation} and \ref{Generation Rate},

\begin{equation}
\begin{split}
\dfrac{\partial f_g}{\partial t}&=\sqrt{\dfrac{\pi}{2}}\dfrac{e^2 v_F^2 w}{2 \hbar^2}A_o^2 \vert \sigma_{cv}^{\lambda}\vert^2 \exp{\left(-\dfrac{t^2}{2 w^2}\right)} \exp{\left(-\dfrac{w^2 \delta^2}{2}\right)}\\
&\times (f_v-f_c) \text{Real}\left[ e^{-i\delta t}\left( 1+\text{Erf}\left( \dfrac{t}{\sqrt{2}w}-\dfrac{i\delta w}{\sqrt{2}}\right)\right)\right]
\end{split}
\end{equation}
where Erf is the error function, $w$ is the pump pulse width, $\delta=\omega_{cv}-\omega_o$ is the detuning. The factor $\vert \sigma^\lambda_{cv}\vert^2$ implies the interband matrix element of the sub-lattice Pauli matrix in the ($k_x,k_y$)-plane for linearly polarized pump pulse with polarization in $\lambda$ direction.\\

The collision integral terms are approximated within the Relaxation Time Approximation (RTA) as described below.
\begin{figure}[htbp]
\centering
\includegraphics[scale=0.35]{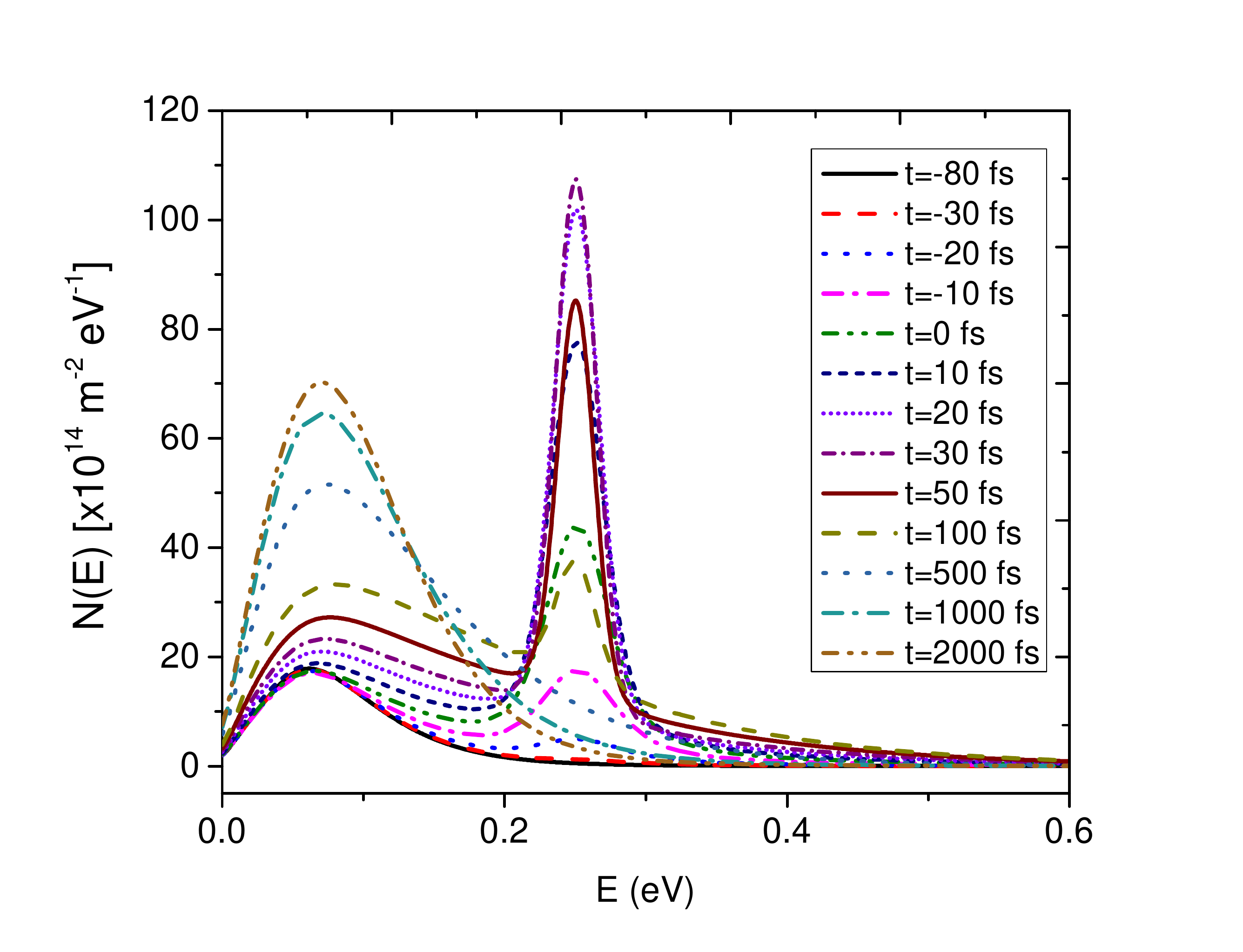}
\caption{\label{energy_contour_0meV}(color online) Time evolution of $N(E)$ [defined in Eq.~\ref{DOStimesF}], the product of the density of states per unit area and the probability of occupation of state with energy $E$ for undoped graphene as a function of energy.}
\end{figure}

The first term in the collision integral corresponds to `Thermalization' where the distribution relaxes  via electron-electron intraband scattering \cite{ZhouWuDriftedFD2010} to a drifted Fermi-Dirac distribution with a non-zero time dependent chemical potential and an effective electron temperature that is higher than the lattice temperature
\begin{equation}
\label{Thermalization}
f_t=[1+\exp\{(\hbar v_F \vert \vec{k}-\vec{k_d}\vert-\mu^{*})/(k_b T_{el})\}]^{-1}.
\end{equation}

\begin{figure*}
\centering
\includegraphics[width=6 in,height=3.5 in]{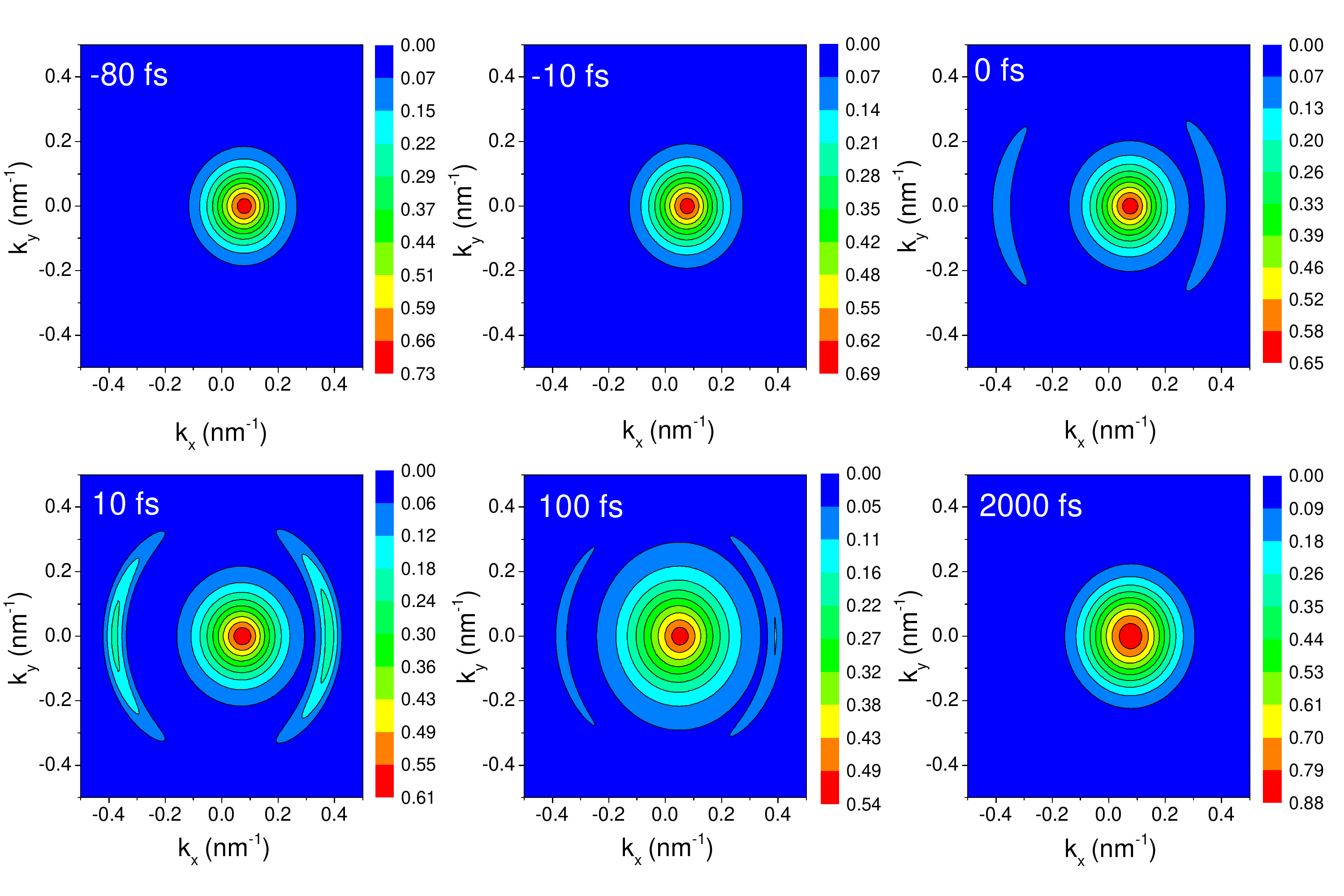}
\caption{\label{contour_50meV}(color online) Time evolution of the distribution function for n-doped graphene with room temperature chemical potential of 50 meV. The carriers are generated by the pump pulse with $Y$-polarization ( $\propto \cos^2 \theta$) during the pulse width duration. The carriers then thermalize to a hot Fermi-Dirac distribution which eventually cools down to the lattice temperature.}
\end{figure*}

\begin{figure}[htbp]
\centering
\includegraphics[scale=0.35]{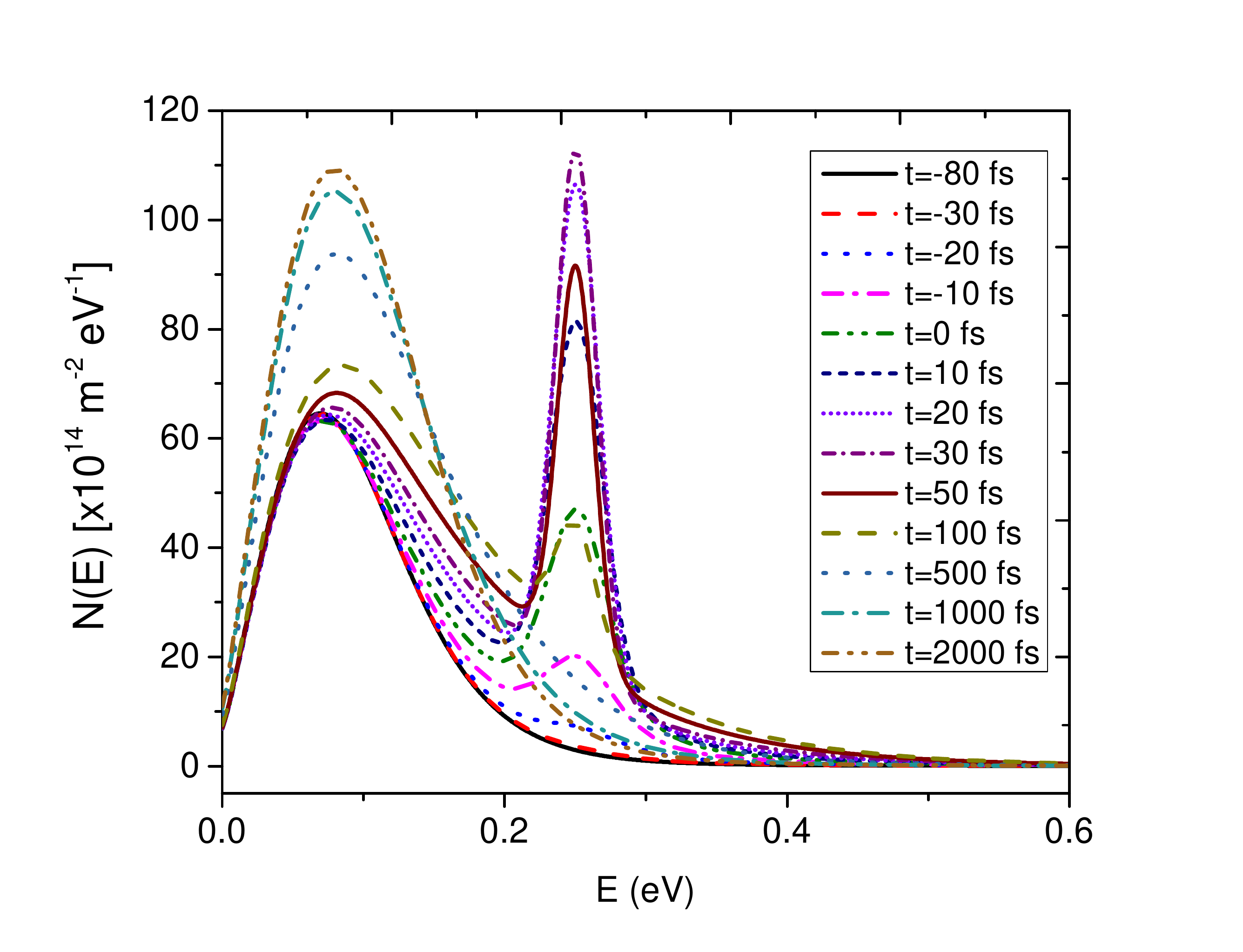}
\caption{\label{energy_contour_50meV}(color online) Time evolution of $N(E)$ [defined in Eq.~\ref{DOStimesF}], the product of the density of states per unit area and the probability of occupation of state with energy $E$ for for n-doped graphene with room temperature chemical potential of 50 meV as a function of energy.}
\end{figure}

We note that the `Thermalization' term in the collision integral is due to scattering between carriers and thus conserves number density, energy and wave-vector. Since the DC field is assumed to be along the x-direction:
\begin{equation}
\begin{split}
&k_d=\dfrac{\langle k_x \rangle}{n}=\dfrac{1}{n}\iint \dfrac{d\vec{k}}{(2\pi)^2} k_x f_t (\vec{k}) \\
&n(t)=\iint \dfrac{d\vec{k}}{(2\pi)^2}  f_t (\vec{k})\\
&\langle \epsilon_k \rangle=\iint \dfrac{d\vec{k}}{(2\pi)^2}  \epsilon_k f_t (\vec{k})
\end{split}
\end{equation}
The equations above determine the quasi-Fermi level ($\mu^*$), electron temperature ($T_{el}$) and the drift wave-vector ($k_d$). The quasi-Fermi level $\mu^*$ and the electron temperature increases over the duration of the pump pulse as the pump generates carriers in high energy states. \\

The second term in the collision integral corresponds to `cooling' of carriers generated by the pump and here the distribution relaxes to a Fermi-Dirac distribution with a non-zero chemical potential and lattice temperature ($T$) via intraband scattering with phonons.
\begin{equation}
\label{Cooling}
f_c=[1+\exp\{(\hbar v_F \vert \vec{k} \vert -\mu)/(k_b T)\}]^{-1}.
\end{equation}
The collision integral term for `Cooling' of carriers also conserves carrier density. This conservation rule determines the Fermi-level ($\mu$) which  increases over the duration of the pump pulse.\\

\subsection{Observables}

The Boltzmann equation can be solved using Fourier transforms as described in Appendix.~\ref{Solution to BTE} to obtain the time dependent distribution function. The carriers are in a steady-state prior to the application of the pump. The time dependent distribution obtained can be used to evaluate the transient current density
\begin{equation}
\vec{j}=-e\iint \dfrac{d\vec{k}}{(2\pi)^2} \vec{v}f(\vec{k},t).
\end{equation}

The radiation for the charge carrier acceleration is proportional to the time derivative of the current.
\begin{equation}
\dfrac{d\vec{j}}{dt}=-e\iint \dfrac{d\vec{k}}{(2\pi)^2} \vec{v}\dfrac{df(\vec{k},t)}{dt}.
\end{equation}


\section{Results}
\label{Results}

Assuming the pump vector potential is of the Gaussian form in Eq.~(\ref{GaussianVectorPotential}). The fluence of the pump pulse `$F$' and the central energy `$\omega_o$' can be used to calculate `$A_o$': $$A_o\approx\sqrt{\dfrac{2F}{\sqrt{\pi}\omega_o^2 \varepsilon_o w c}}.$$ Also given the FWHM (full width half maximum) of the pump `$p_d$', the parameter `$w$ =$p_d/(2\sqrt{2\text{ln}(2)})$' is evaluated. The parameters chosen for the calculations are presented in Table.~\ref{parameters_THz}.\\

\begin{table}
\caption{\label{parameters_THz} Parameters used for calculating THz radiation contribution from photoexcited carrier acceleration. }
\begin{ruledtabular}
\begin{tabular}{lr}
Parameter & Value \\
\hline
Fluence $F $	  &1 $\mu$J/cm$^2$\\
Pulse Central Energy $\hbar\omega_o$ & 0.5 eV\\
FWHM $p_d$ & 40 fs \\
Thermalization time $\tau_T$ & 50 fs\\
Cooling time $\tau_C$ & 0.5 ps \\
Electric Field $E$ & 1 kV/cm \\
Temperature $T$ & 300 K\\
\end{tabular}
\end{ruledtabular}
\end{table}

The results presented below are for undoped graphene as well as for n-doped graphene with for n-doped graphene with room temperature chemical potential of 50 meV and Y-polarized pump pulse.\\

\begin{figure}[htbp]
\centering
\includegraphics[scale=0.35]{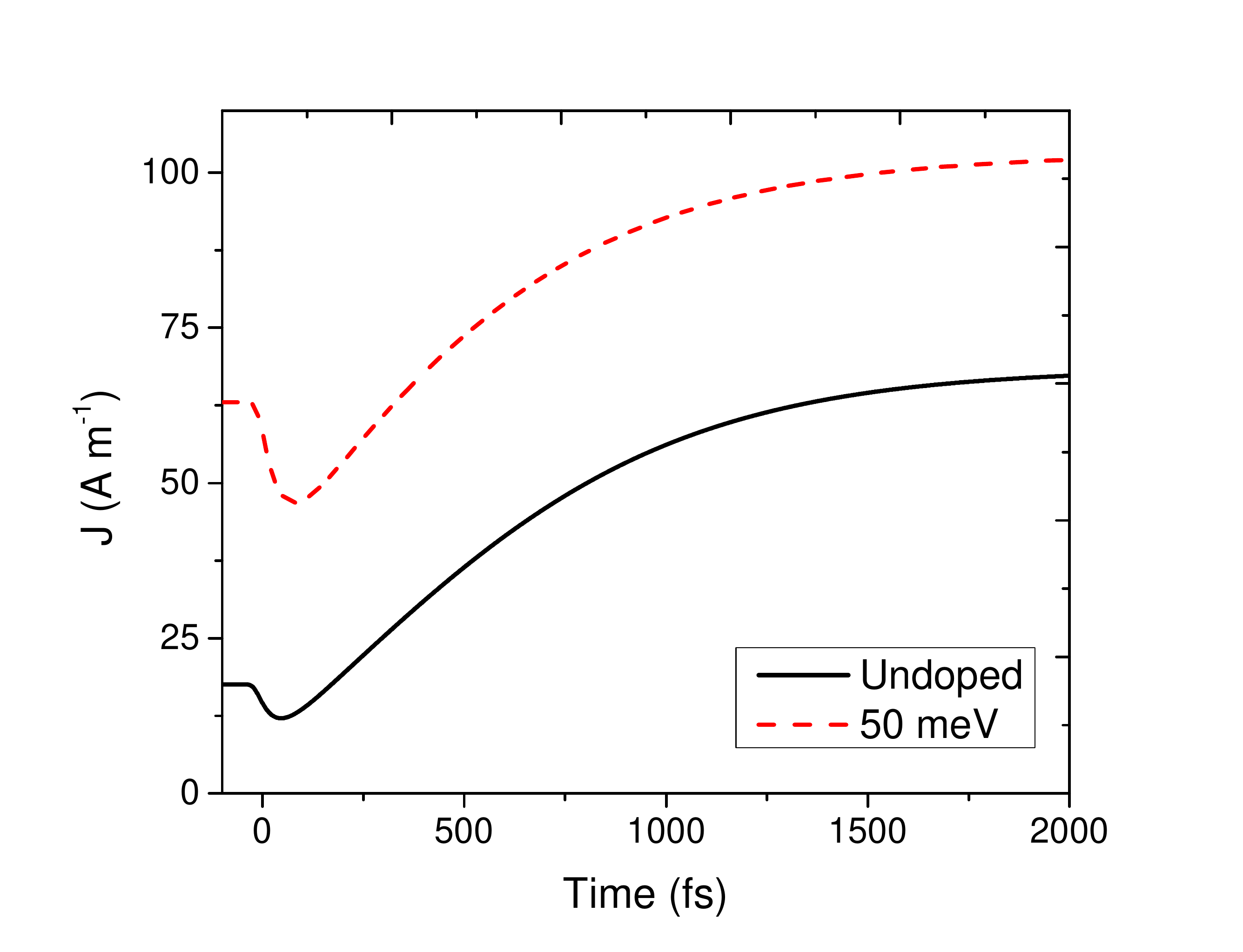}
\caption{\label{current}(color online) Total current density of all carriers as a function of time for undoped and n-doped graphene with room temperature chemical potential of 50 meV.}
\end{figure}
\begin{figure}[htbp]
\centering
\includegraphics[scale=0.35]{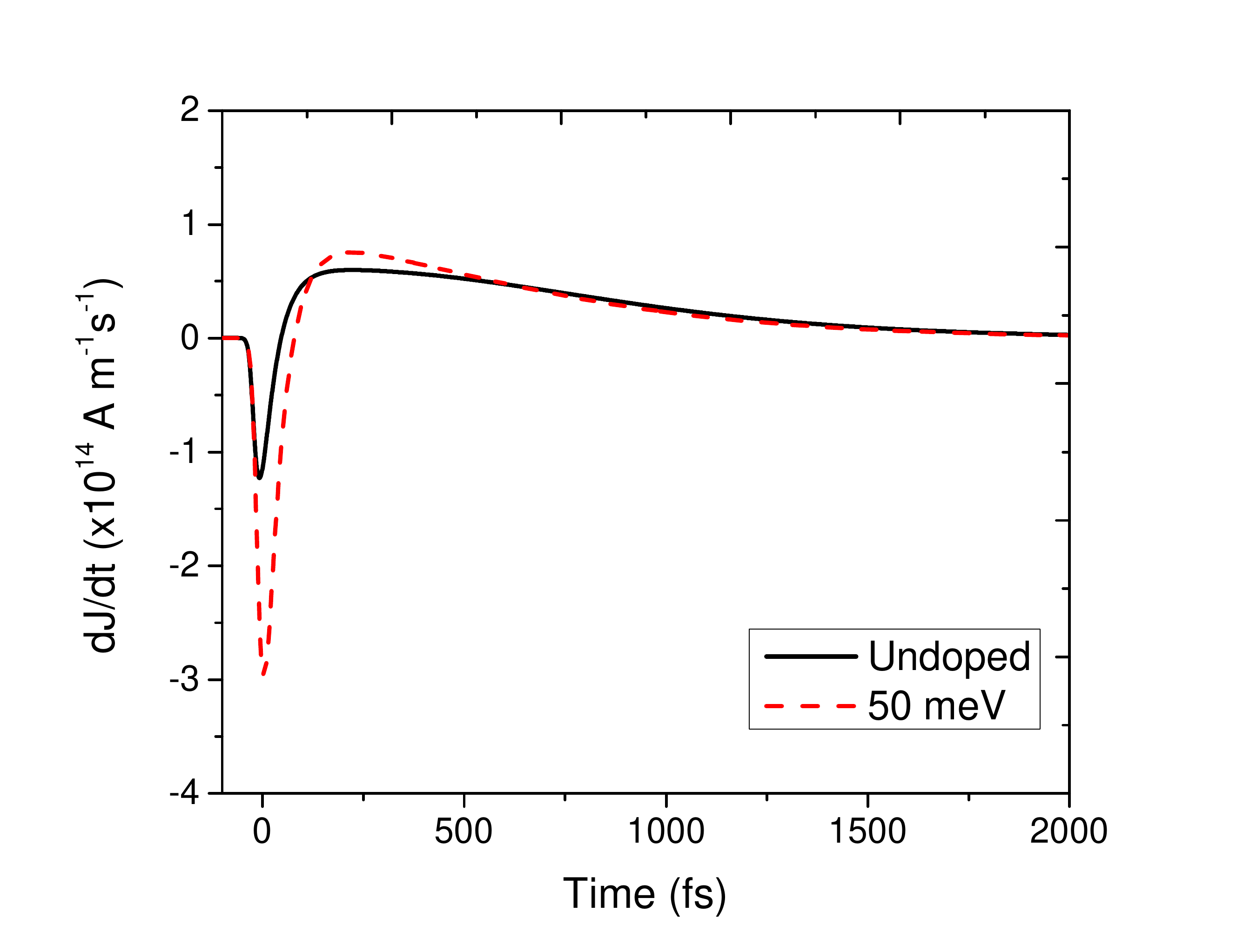}
\caption{\label{tderivative_current}(color online) Time derivative of total current density of all carriers as a function of time for undoped and n-doped graphene with room temperature chemical potential of 50 meV.}
\end{figure}

The time evolution of the distribution function for undoped graphene as shown in Fig.~(\ref{contour_0meV}) and n-doped graphene with room temperature chemical potential of 50 meV in Fig.~(\ref{contour_50meV}). The carriers are generated anisotropically by the polarized pump pulse \cite{MalicAbsorptionGraphene2011} after which the thermalization of carriers takes place and subsequently the hot carriers cool down by interaction with the lattice. The distribution of carriers in energy space is shown for different times ranging from -80 fs to 2 ps for undoped graphene in Fig.~(\ref{energy_contour_0meV}) and n-doped graphene with room temperature chemical potential of 50 meV in Fig.~(\ref{energy_contour_50meV}) respectively.\\

The distribution of carrier density in energy $N(E)$ can be evaluated. $N(E)dE$ is the carrier density between the energies $E$ and $E+dE$ which when integrated over all energies gives the total number density. This energy distribution is evaluated using
\begin{equation}
\label{DOStimesF}
N(E)=\iint \dfrac{d\bm{k}}{(2\pi)^2}\, f(\bm{k}) \delta\left(\varepsilon_{\bm{k}}-E\right).
\end{equation}
This is equivalent to the product of the density of states per unit area and the probability of occupation of state with energy $E$. The distribution of carriers in energy space is shown in Fig.~\ref{energy_contour_0meV} and Fig.~\ref{energy_contour_50meV}. It shows the generation of carriers by the optical pump pulse during the pulse duration. The carriers then thermalize and eventually cool on longer time scales.\\

The current density averaged over the distribution shown in Fig.~\ref{current} initially decreases during the generation of carriers. This is because the sudden increase in carrier density created by the pump pulse actually \textit{decreases} the drift wavevector (see Fig.~\ref{drift_wavevector}) and the fast thermalization relaxes the system towards a distribution is centered about a smaller drift wavevector $k_d$ causing a \textit{decrease} in the average velocity/current. At longer time scales the averaged current density increases as the carrier distribution drifts and eventually saturates over the cooling time scale.

The time derivative of the averaged current density over the distribution is also plotted in Fig.~\ref{tderivative_current}. It decreases over the pumping duration and increases when carriers drift before decaying to zero when the carriers cool down and reach a steady-state.

\section{Conclusion}
The radiative contribution to the THz signal of ultrafast photoexcited carrier acceleration in presence of an in-plane DC electric field in graphene is studied. The linearity of the graphene dispersion near the Dirac point implies constant magnitude of velocity which naively might lead one to expect no THz radiation. However the azimuthal degree of freedom allows for a time dependent velocity/current density where the \textit{direction} of the current is changing. The polarized pump pulse creates an anisotropic carrier distribution in the 2D Brillouin Zone. Since the thermalization time scale arising from rapid carrier-carrier scattering is of the order of 50 fs, the anisotropic photoexcited carrier distribution relaxes to a drifted Fermi-Dirac distribution at a higher temperature relative to the lattice. The cooling of lattice through phonons happens on a time scale of ps. It is this cooling which via momentum and energy relaxation drives the current density to its steady-state value. Thus the two relaxation mechanisms give rise to a time-varying current density which radiates in the THz frequency range. In spite of a constant ``speed" of the carriers in graphene, there is still radiation from the acceleration of the carriers as they change their \textit{direction}.

\section{Acknowledgments}
This work was supported by AFOSR through grant FA9550-14-1-0376.
\bibliography{Graphene_THz_ref}

\begin{thebibliography}{29}%
\makeatletter
\providecommand \@ifxundefined [1]{%
 \@ifx{#1\undefined}
}%
\providecommand \@ifnum [1]{%
 \ifnum #1\expandafter \@firstoftwo
 \else \expandafter \@secondoftwo
 \fi
}%
\providecommand \@ifx [1]{%
 \ifx #1\expandafter \@firstoftwo
 \else \expandafter \@secondoftwo
 \fi
}%
\providecommand \natexlab [1]{#1}%
\providecommand \enquote  [1]{``#1''}%
\providecommand \bibnamefont  [1]{#1}%
\providecommand \bibfnamefont [1]{#1}%
\providecommand \citenamefont [1]{#1}%
\providecommand \href@noop [0]{\@secondoftwo}%
\providecommand \href [0]{\begingroup \@sanitize@url \@href}%
\providecommand \@href[1]{\@@startlink{#1}\@@href}%
\providecommand \@@href[1]{\endgroup#1\@@endlink}%
\providecommand \@sanitize@url [0]{\catcode `\\12\catcode `\$12\catcode
  `\&12\catcode `\#12\catcode `\^12\catcode `\_12\catcode `\%12\relax}%
\providecommand \@@startlink[1]{}%
\providecommand \@@endlink[0]{}%
\providecommand \url  [0]{\begingroup\@sanitize@url \@url }%
\providecommand \@url [1]{\endgroup\@href {#1}{\urlprefix }}%
\providecommand \urlprefix  [0]{URL }%
\providecommand \Eprint [0]{\href }%
\providecommand \doibase [0]{http://dx.doi.org/}%
\providecommand \selectlanguage [0]{\@gobble}%
\providecommand \bibinfo  [0]{\@secondoftwo}%
\providecommand \bibfield  [0]{\@secondoftwo}%
\providecommand \translation [1]{[#1]}%
\providecommand \BibitemOpen [0]{}%
\providecommand \bibitemStop [0]{}%
\providecommand \bibitemNoStop [0]{.\EOS\space}%
\providecommand \EOS [0]{\spacefactor3000\relax}%
\providecommand \BibitemShut  [1]{\csname bibitem#1\endcsname}%
\let\auto@bib@innerbib\@empty
\bibitem [{\citenamefont {De~Maagt}\ \emph {et~al.}(2005)\citenamefont
  {De~Maagt}, \citenamefont {Bolivar},\ and\ \citenamefont
  {Mann}}]{Maagt_THzBook}%
  \BibitemOpen
  \bibfield  {author} {\bibinfo {author} {\bibfnamefont {P.}~\bibnamefont
  {De~Maagt}}, \bibinfo {author} {\bibfnamefont {P.~H.}\ \bibnamefont
  {Bolivar}}, \ and\ \bibinfo {author} {\bibfnamefont {C.}~\bibnamefont
  {Mann}},\ }\href@noop {} {\bibfield  {journal} {\bibinfo  {journal}
  {Encyclopedia of RF and Microwave Engineering}\ } (\bibinfo {year}
  {2005})}\BibitemShut {NoStop}%
\bibitem [{\citenamefont {Tan}\ \emph {et~al.}(2012)\citenamefont {Tan},
  \citenamefont {Huang}, \citenamefont {Liu}, \citenamefont {Xiong},\ and\
  \citenamefont {Fan}}]{Tan2012FEL}%
  \BibitemOpen
  \bibfield  {author} {\bibinfo {author} {\bibfnamefont {P.}~\bibnamefont
  {Tan}}, \bibinfo {author} {\bibfnamefont {J.}~\bibnamefont {Huang}}, \bibinfo
  {author} {\bibfnamefont {K.}~\bibnamefont {Liu}}, \bibinfo {author}
  {\bibfnamefont {Y.}~\bibnamefont {Xiong}}, \ and\ \bibinfo {author}
  {\bibfnamefont {M.}~\bibnamefont {Fan}},\ }\href {\doibase
  10.1007/s11432-011-4515-1} {\bibfield  {journal} {\bibinfo  {journal}
  {Science China Information Sciences}\ }\textbf {\bibinfo {volume} {55}},\
  \bibinfo {pages} {1} (\bibinfo {year} {2012})}\BibitemShut {NoStop}%
\bibitem [{\citenamefont {Williams}(2007)}]{THzQCL}%
  \BibitemOpen
  \bibfield  {author} {\bibinfo {author} {\bibfnamefont {B.~S.}\ \bibnamefont
  {Williams}},\ }\href {\doibase 10.1038/nphoton.2007.166} {\bibfield
  {journal} {\bibinfo  {journal} {Nature Photonics}\ }\textbf {\bibinfo
  {volume} {1}},\ \bibinfo {pages} {517} (\bibinfo {year} {2007})}\BibitemShut
  {NoStop}%
\bibitem [{\citenamefont {Alekseev}\ and\ \citenamefont
  {Pavlidis}(2000)}]{alekseev2000gan}%
  \BibitemOpen
  \bibfield  {author} {\bibinfo {author} {\bibfnamefont {E.}~\bibnamefont
  {Alekseev}}\ and\ \bibinfo {author} {\bibfnamefont {D.}~\bibnamefont
  {Pavlidis}},\ }in\ \href@noop {} {\emph {\bibinfo {booktitle} {IEEE MTT-S
  International Microwave Symposium Digest}}},\ Vol.~\bibinfo {volume} {3}\
  (\bibinfo {year} {2000})\ pp.\ \bibinfo {pages} {1905--1908}\BibitemShut
  {NoStop}%
\bibitem [{\citenamefont {Pala}\ and\ \citenamefont
  {Abbas}(2012)}]{pala2012terahertz}%
  \BibitemOpen
  \bibfield  {author} {\bibinfo {author} {\bibfnamefont {N.}~\bibnamefont
  {Pala}}\ and\ \bibinfo {author} {\bibfnamefont {A.~N.}\ \bibnamefont
  {Abbas}},\ }\href@noop {} {\bibfield  {journal} {\bibinfo  {journal}
  {Encyclopedia of Nanotechnology}\ ,\ \bibinfo {pages} {2653}} (\bibinfo
  {year} {2012})}\BibitemShut {NoStop}%
\bibitem [{\citenamefont {Tonouchi}(2007)}]{Tonouchi2007}%
  \BibitemOpen
  \bibfield  {author} {\bibinfo {author} {\bibfnamefont {M.}~\bibnamefont
  {Tonouchi}},\ }\href {\doibase 10.1038/nphoton.2007.3} {\bibfield  {journal}
  {\bibinfo  {journal} {Nature Photonics}\ }\textbf {\bibinfo {volume} {1}},\
  \bibinfo {pages} {97} (\bibinfo {year} {2007})}\BibitemShut {NoStop}%
\bibitem [{\citenamefont {Ma}\ and\ \citenamefont
  {Zhang}(1993)}]{MaAndZhang1993}%
  \BibitemOpen
  \bibfield  {author} {\bibinfo {author} {\bibfnamefont {X.}~\bibnamefont
  {Ma}}\ and\ \bibinfo {author} {\bibfnamefont {X.}~\bibnamefont {Zhang}},\
  }\href {\doibase 10.1364/JOSAB.10.001175} {\bibfield  {journal} {\bibinfo
  {journal} {J. Opt. Soc. Am. B}\ }\textbf {\bibinfo {volume} {10}},\ \bibinfo
  {pages} {1175} (\bibinfo {year} {1993})}\BibitemShut {NoStop}%
\bibitem [{\citenamefont {Rice}\ \emph {et~al.}(1994)\citenamefont {Rice},
  \citenamefont {Jin}, \citenamefont {Ma}, \citenamefont {Zhang}, \citenamefont
  {Bliss}, \citenamefont {Larkin},\ and\ \citenamefont
  {Alexander}}]{RiceMaAndZhang1994}%
  \BibitemOpen
  \bibfield  {author} {\bibinfo {author} {\bibfnamefont {A.}~\bibnamefont
  {Rice}}, \bibinfo {author} {\bibfnamefont {Y.}~\bibnamefont {Jin}}, \bibinfo
  {author} {\bibfnamefont {X.}~\bibnamefont {Ma}}, \bibinfo {author}
  {\bibfnamefont {X.}~\bibnamefont {Zhang}}, \bibinfo {author} {\bibfnamefont
  {D.}~\bibnamefont {Bliss}}, \bibinfo {author} {\bibfnamefont
  {J.}~\bibnamefont {Larkin}}, \ and\ \bibinfo {author} {\bibfnamefont
  {M.}~\bibnamefont {Alexander}},\ }\href {\doibase
  http://dx.doi.org/10.1063/1.111922} {\bibfield  {journal} {\bibinfo
  {journal} {Appl. Phys. Lett.}\ }\textbf {\bibinfo {volume} {64}},\ \bibinfo
  {pages} {1324} (\bibinfo {year} {1994})}\BibitemShut {NoStop}%
\bibitem [{\citenamefont {Kuznetsov}\ and\ \citenamefont
  {Stanton}(1993)}]{KuznetsovStanton1993}%
  \BibitemOpen
  \bibfield  {author} {\bibinfo {author} {\bibfnamefont {A.~V.}\ \bibnamefont
  {Kuznetsov}}\ and\ \bibinfo {author} {\bibfnamefont {C.~J.}\ \bibnamefont
  {Stanton}},\ }\href {\doibase 10.1103/PhysRevB.48.10828} {\bibfield
  {journal} {\bibinfo  {journal} {Phys. Rev. B}\ }\textbf {\bibinfo {volume}
  {48}},\ \bibinfo {pages} {10828} (\bibinfo {year} {1993})}\BibitemShut
  {NoStop}%
\bibitem [{\citenamefont {Hu}\ \emph {et~al.}(1994)\citenamefont {Hu},
  \citenamefont {Weling}, \citenamefont {Auston}, \citenamefont {Kuznetsov},\
  and\ \citenamefont {Stanton}}]{KuznetsovStantonAuston1994}%
  \BibitemOpen
  \bibfield  {author} {\bibinfo {author} {\bibfnamefont {B.~B.}\ \bibnamefont
  {Hu}}, \bibinfo {author} {\bibfnamefont {A.~S.}\ \bibnamefont {Weling}},
  \bibinfo {author} {\bibfnamefont {D.~H.}\ \bibnamefont {Auston}}, \bibinfo
  {author} {\bibfnamefont {A.~V.}\ \bibnamefont {Kuznetsov}}, \ and\ \bibinfo
  {author} {\bibfnamefont {C.~J.}\ \bibnamefont {Stanton}},\ }\href {\doibase
  10.1103/PhysRevB.49.2234} {\bibfield  {journal} {\bibinfo  {journal} {Phys.
  Rev. B}\ }\textbf {\bibinfo {volume} {49}},\ \bibinfo {pages} {2234}
  (\bibinfo {year} {1994})}\BibitemShut {NoStop}%
\bibitem [{\citenamefont {Dekorsy}\ \emph {et~al.}(1996)\citenamefont
  {Dekorsy}, \citenamefont {Auer}, \citenamefont {Bakker}, \citenamefont
  {Roskos},\ and\ \citenamefont {Kurz}}]{Dekorsy1996}%
  \BibitemOpen
  \bibfield  {author} {\bibinfo {author} {\bibfnamefont {T.}~\bibnamefont
  {Dekorsy}}, \bibinfo {author} {\bibfnamefont {H.}~\bibnamefont {Auer}},
  \bibinfo {author} {\bibfnamefont {H.~J.}\ \bibnamefont {Bakker}}, \bibinfo
  {author} {\bibfnamefont {H.~G.}\ \bibnamefont {Roskos}}, \ and\ \bibinfo
  {author} {\bibfnamefont {H.}~\bibnamefont {Kurz}},\ }\href {\doibase
  10.1103/PhysRevB.53.4005} {\bibfield  {journal} {\bibinfo  {journal} {Phys.
  Rev. B}\ }\textbf {\bibinfo {volume} {53}},\ \bibinfo {pages} {4005}
  (\bibinfo {year} {1996})}\BibitemShut {NoStop}%
\bibitem [{\citenamefont {Johnston}\ \emph {et~al.}(2002)\citenamefont
  {Johnston}, \citenamefont {Whittaker}, \citenamefont {Corchia}, \citenamefont
  {Davies},\ and\ \citenamefont {Linfield}}]{Johnston2002}%
  \BibitemOpen
  \bibfield  {author} {\bibinfo {author} {\bibfnamefont {M.~B.}\ \bibnamefont
  {Johnston}}, \bibinfo {author} {\bibfnamefont {D.~M.}\ \bibnamefont
  {Whittaker}}, \bibinfo {author} {\bibfnamefont {A.}~\bibnamefont {Corchia}},
  \bibinfo {author} {\bibfnamefont {A.~G.}\ \bibnamefont {Davies}}, \ and\
  \bibinfo {author} {\bibfnamefont {E.~H.}\ \bibnamefont {Linfield}},\ }\href
  {\doibase 10.1103/PhysRevB.65.165301} {\bibfield  {journal} {\bibinfo
  {journal} {Phys. Rev. B}\ }\textbf {\bibinfo {volume} {65}},\ \bibinfo
  {pages} {165301} (\bibinfo {year} {2002})}\BibitemShut {NoStop}%
\bibitem [{\citenamefont {Zhang}\ and\ \citenamefont
  {Auston}(1992)}]{ZhangAuston1992}%
  \BibitemOpen
  \bibfield  {author} {\bibinfo {author} {\bibfnamefont {X.}~\bibnamefont
  {Zhang}}\ and\ \bibinfo {author} {\bibfnamefont {D.}~\bibnamefont {Auston}},\
  }\href {\doibase http://dx.doi.org/10.1063/1.350710} {\bibfield  {journal}
  {\bibinfo  {journal} {J. Appl. Phys.}\ }\textbf {\bibinfo {volume} {71}},\
  \bibinfo {pages} {326} (\bibinfo {year} {1992})}\BibitemShut {NoStop}%
\bibitem [{\citenamefont {Dekorsy}\ \emph {et~al.}(1993)\citenamefont
  {Dekorsy}, \citenamefont {Pfeifer}, \citenamefont {K\"utt},\ and\
  \citenamefont {Kurz}}]{Dekorsy1993}%
  \BibitemOpen
  \bibfield  {author} {\bibinfo {author} {\bibfnamefont {T.}~\bibnamefont
  {Dekorsy}}, \bibinfo {author} {\bibfnamefont {T.}~\bibnamefont {Pfeifer}},
  \bibinfo {author} {\bibfnamefont {W.}~\bibnamefont {K\"utt}}, \ and\ \bibinfo
  {author} {\bibfnamefont {H.}~\bibnamefont {Kurz}},\ }\href {\doibase
  10.1103/PhysRevB.47.3842} {\bibfield  {journal} {\bibinfo  {journal} {Phys.
  Rev. B}\ }\textbf {\bibinfo {volume} {47}},\ \bibinfo {pages} {3842}
  (\bibinfo {year} {1993})}\BibitemShut {NoStop}%
\bibitem [{\citenamefont {Novoselov}\ \emph {et~al.}(2005)\citenamefont
  {Novoselov}, \citenamefont {Geim}, \citenamefont {Morozov}, \citenamefont
  {Jiang}, \citenamefont {Grigorieva}, \citenamefont {Dubonos},\ and\
  \citenamefont {Firsov}}]{novoselov2005two}%
  \BibitemOpen
  \bibfield  {author} {\bibinfo {author} {\bibfnamefont {K.}~\bibnamefont
  {Novoselov}}, \bibinfo {author} {\bibfnamefont {A.~K.}\ \bibnamefont {Geim}},
  \bibinfo {author} {\bibfnamefont {S.}~\bibnamefont {Morozov}}, \bibinfo
  {author} {\bibfnamefont {D.}~\bibnamefont {Jiang}}, \bibinfo {author}
  {\bibfnamefont {M.~K.~I.}\ \bibnamefont {Grigorieva}}, \bibinfo {author}
  {\bibfnamefont {S.}~\bibnamefont {Dubonos}}, \ and\ \bibinfo {author}
  {\bibfnamefont {A.}~\bibnamefont {Firsov}},\ }\href {\doibase
  10.1038/nature04233} {\bibfield  {journal} {\bibinfo  {journal} {Nature}\
  }\textbf {\bibinfo {volume} {438}},\ \bibinfo {pages} {197} (\bibinfo {year}
  {2005})}\BibitemShut {NoStop}%
\bibitem [{\citenamefont {Booshehri}\ \emph {et~al.}(2012)\citenamefont
  {Booshehri}, \citenamefont {Mielke}, \citenamefont {Rickel}, \citenamefont
  {Crooker}, \citenamefont {Zhang}, \citenamefont {Ren}, \citenamefont
  {H\'aroz}, \citenamefont {Rustagi}, \citenamefont {Stanton}, \citenamefont
  {Jin}, \citenamefont {Sun}, \citenamefont {Yan}, \citenamefont {Tour},\ and\
  \citenamefont {Kono}}]{Booshehri:2012CR}%
  \BibitemOpen
  \bibfield  {author} {\bibinfo {author} {\bibfnamefont {L.~G.}\ \bibnamefont
  {Booshehri}}, \bibinfo {author} {\bibfnamefont {C.~H.}\ \bibnamefont
  {Mielke}}, \bibinfo {author} {\bibfnamefont {D.~G.}\ \bibnamefont {Rickel}},
  \bibinfo {author} {\bibfnamefont {S.~A.}\ \bibnamefont {Crooker}}, \bibinfo
  {author} {\bibfnamefont {Q.}~\bibnamefont {Zhang}}, \bibinfo {author}
  {\bibfnamefont {L.}~\bibnamefont {Ren}}, \bibinfo {author} {\bibfnamefont
  {E.~H.}\ \bibnamefont {H\'aroz}}, \bibinfo {author} {\bibfnamefont
  {A.}~\bibnamefont {Rustagi}}, \bibinfo {author} {\bibfnamefont {C.~J.}\
  \bibnamefont {Stanton}}, \bibinfo {author} {\bibfnamefont {Z.}~\bibnamefont
  {Jin}}, \bibinfo {author} {\bibfnamefont {Z.}~\bibnamefont {Sun}}, \bibinfo
  {author} {\bibfnamefont {Z.}~\bibnamefont {Yan}}, \bibinfo {author}
  {\bibfnamefont {J.~M.}\ \bibnamefont {Tour}}, \ and\ \bibinfo {author}
  {\bibfnamefont {J.}~\bibnamefont {Kono}},\ }\href {\doibase
  10.1103/PhysRevB.85.205407} {\bibfield  {journal} {\bibinfo  {journal} {Phys.
  Rev. B}\ }\textbf {\bibinfo {volume} {85}},\ \bibinfo {pages} {205407}
  (\bibinfo {year} {2012})}\BibitemShut {NoStop}%
\bibitem [{\citenamefont {Gusynin}\ and\ \citenamefont
  {Sharapov}(2005)}]{Guysnin2005_QHE}%
  \BibitemOpen
  \bibfield  {author} {\bibinfo {author} {\bibfnamefont {V.~P.}\ \bibnamefont
  {Gusynin}}\ and\ \bibinfo {author} {\bibfnamefont {S.~G.}\ \bibnamefont
  {Sharapov}},\ }\href {\doibase 10.1103/PhysRevLett.95.146801} {\bibfield
  {journal} {\bibinfo  {journal} {Phys. Rev. Lett.}\ }\textbf {\bibinfo
  {volume} {95}},\ \bibinfo {pages} {146801} (\bibinfo {year}
  {2005})}\BibitemShut {NoStop}%
\bibitem [{\citenamefont {Zhang}\ \emph {et~al.}(2005)\citenamefont {Zhang},
  \citenamefont {Tan}, \citenamefont {Stormer},\ and\ \citenamefont
  {Kim}}]{Zhang2005_QHE}%
  \BibitemOpen
  \bibfield  {author} {\bibinfo {author} {\bibfnamefont {Y.}~\bibnamefont
  {Zhang}}, \bibinfo {author} {\bibfnamefont {Y.}~\bibnamefont {Tan}}, \bibinfo
  {author} {\bibfnamefont {H.}~\bibnamefont {Stormer}}, \ and\ \bibinfo
  {author} {\bibfnamefont {P.}~\bibnamefont {Kim}},\ }\href {\doibase
  10.1038/nature04235} {\bibfield  {journal} {\bibinfo  {journal} {Nature}\
  }\textbf {\bibinfo {volume} {438}},\ \bibinfo {pages} {201} (\bibinfo {year}
  {2005})}\BibitemShut {NoStop}%
\bibitem [{\citenamefont {Rustagi}\ and\ \citenamefont
  {Stanton}(2014)}]{Rustagi2014}%
  \BibitemOpen
  \bibfield  {author} {\bibinfo {author} {\bibfnamefont {A.}~\bibnamefont
  {Rustagi}}\ and\ \bibinfo {author} {\bibfnamefont {C.~J.}\ \bibnamefont
  {Stanton}},\ }\href {\doibase 10.1103/PhysRevB.90.245424} {\bibfield
  {journal} {\bibinfo  {journal} {Phys. Rev. B}\ }\textbf {\bibinfo {volume}
  {90}},\ \bibinfo {pages} {245424} (\bibinfo {year} {2014})}\BibitemShut
  {NoStop}%
\bibitem [{\citenamefont {Castro~Neto}\ \emph {et~al.}(2009)\citenamefont
  {Castro~Neto}, \citenamefont {Guinea}, \citenamefont {Peres}, \citenamefont
  {Novoselov},\ and\ \citenamefont {Geim}}]{CastroNetoRMP}%
  \BibitemOpen
  \bibfield  {author} {\bibinfo {author} {\bibfnamefont {A.~H.}\ \bibnamefont
  {Castro~Neto}}, \bibinfo {author} {\bibfnamefont {F.}~\bibnamefont {Guinea}},
  \bibinfo {author} {\bibfnamefont {N.~M.~R.}\ \bibnamefont {Peres}}, \bibinfo
  {author} {\bibfnamefont {K.~S.}\ \bibnamefont {Novoselov}}, \ and\ \bibinfo
  {author} {\bibfnamefont {A.~K.}\ \bibnamefont {Geim}},\ }\href {\doibase
  10.1103/RevModPhys.81.109} {\bibfield  {journal} {\bibinfo  {journal} {Rev.
  Mod. Phys.}\ }\textbf {\bibinfo {volume} {81}},\ \bibinfo {pages} {109}
  (\bibinfo {year} {2009})}\BibitemShut {NoStop}%
\bibitem [{\citenamefont {Glazov}\ and\ \citenamefont
  {Ganichev}(2014)}]{GlazovGanichev2014}%
  \BibitemOpen
  \bibfield  {author} {\bibinfo {author} {\bibfnamefont {M.~M.}\ \bibnamefont
  {Glazov}}\ and\ \bibinfo {author} {\bibfnamefont {S.~D.}\ \bibnamefont
  {Ganichev}},\ }\href {\doibase
  http://dx.doi.org/10.1016/j.physrep.2013.10.003} {\bibfield  {journal}
  {\bibinfo  {journal} {Phys. Rep.}\ }\textbf {\bibinfo {volume} {535}},\
  \bibinfo {pages} {101} (\bibinfo {year} {2014})}\BibitemShut {NoStop}%
\bibitem [{\citenamefont {Maysonnave}\ \emph {et~al.}(2014)\citenamefont
  {Maysonnave}, \citenamefont {Huppert}, \citenamefont {Wang}, \citenamefont
  {Maero}, \citenamefont {Berger}, \citenamefont {de~Heer}, \citenamefont
  {Norris}, \citenamefont {De~Vaulchier}, \citenamefont {Dhillon},
  \citenamefont {Tignon}, \citenamefont {Ferreira},\ and\ \citenamefont
  {Mangeney}}]{Mangeney2014}%
  \BibitemOpen
  \bibfield  {author} {\bibinfo {author} {\bibfnamefont {J.}~\bibnamefont
  {Maysonnave}}, \bibinfo {author} {\bibfnamefont {S.}~\bibnamefont {Huppert}},
  \bibinfo {author} {\bibfnamefont {F.}~\bibnamefont {Wang}}, \bibinfo {author}
  {\bibfnamefont {S.}~\bibnamefont {Maero}}, \bibinfo {author} {\bibfnamefont
  {C.}~\bibnamefont {Berger}}, \bibinfo {author} {\bibfnamefont
  {W.}~\bibnamefont {de~Heer}}, \bibinfo {author} {\bibfnamefont {T.~B.}\
  \bibnamefont {Norris}}, \bibinfo {author} {\bibfnamefont {L.~A.}\
  \bibnamefont {De~Vaulchier}}, \bibinfo {author} {\bibfnamefont
  {S.}~\bibnamefont {Dhillon}}, \bibinfo {author} {\bibfnamefont
  {J.}~\bibnamefont {Tignon}}, \bibinfo {author} {\bibfnamefont
  {R.}~\bibnamefont {Ferreira}}, \ and\ \bibinfo {author} {\bibfnamefont
  {J.}~\bibnamefont {Mangeney}},\ }\href {\doibase 10.1021/nl502684j}
  {\bibfield  {journal} {\bibinfo  {journal} {Nano Lett.}\ }\textbf {\bibinfo
  {volume} {14}},\ \bibinfo {pages} {5797} (\bibinfo {year}
  {2014})}\BibitemShut {NoStop}%
\bibitem [{\citenamefont {Novoselov}\ \emph {et~al.}(2012)\citenamefont
  {Novoselov}, \citenamefont {Fal'ko}, \citenamefont {Colombo}, \citenamefont
  {Gellert}, \citenamefont {Schwab},\ and\ \citenamefont
  {Kim}}]{Graphene_Dember}%
  \BibitemOpen
  \bibfield  {author} {\bibinfo {author} {\bibfnamefont {K.~S.}\ \bibnamefont
  {Novoselov}}, \bibinfo {author} {\bibfnamefont {V.~I.}\ \bibnamefont
  {Fal'ko}}, \bibinfo {author} {\bibfnamefont {L.}~\bibnamefont {Colombo}},
  \bibinfo {author} {\bibfnamefont {P.~R.}\ \bibnamefont {Gellert}}, \bibinfo
  {author} {\bibfnamefont {M.~G.}\ \bibnamefont {Schwab}}, \ and\ \bibinfo
  {author} {\bibfnamefont {K.}~\bibnamefont {Kim}},\ }\href {\doibase
  10.1038/nature11458} {\bibfield  {journal} {\bibinfo  {journal} {Nature}\
  }\textbf {\bibinfo {volume} {490}},\ \bibinfo {pages} {192} (\bibinfo {year}
  {2012})}\BibitemShut {NoStop}%
\bibitem [{\citenamefont {Zhou}\ and\ \citenamefont
  {Wu}(2010)}]{ZhouWuDriftedFD2010}%
  \BibitemOpen
  \bibfield  {author} {\bibinfo {author} {\bibfnamefont {Y.}~\bibnamefont
  {Zhou}}\ and\ \bibinfo {author} {\bibfnamefont {M.~W.}\ \bibnamefont {Wu}},\
  }\href {\doibase 10.1103/PhysRevB.82.085304} {\bibfield  {journal} {\bibinfo
  {journal} {Phys. Rev. B}\ }\textbf {\bibinfo {volume} {82}},\ \bibinfo
  {pages} {085304} (\bibinfo {year} {2010})}\BibitemShut {NoStop}%
\bibitem [{\citenamefont {Malic}\ \emph {et~al.}(2011)\citenamefont {Malic},
  \citenamefont {Winzer}, \citenamefont {Bobkin},\ and\ \citenamefont
  {Knorr}}]{MalicAbsorptionGraphene2011}%
  \BibitemOpen
  \bibfield  {author} {\bibinfo {author} {\bibfnamefont {E.}~\bibnamefont
  {Malic}}, \bibinfo {author} {\bibfnamefont {T.}~\bibnamefont {Winzer}},
  \bibinfo {author} {\bibfnamefont {E.}~\bibnamefont {Bobkin}}, \ and\ \bibinfo
  {author} {\bibfnamefont {A.}~\bibnamefont {Knorr}},\ }\href {\doibase
  10.1103/PhysRevB.84.205406} {\bibfield  {journal} {\bibinfo  {journal} {Phys.
  Rev. B}\ }\textbf {\bibinfo {volume} {84}},\ \bibinfo {pages} {205406}
  (\bibinfo {year} {2011})}\BibitemShut {NoStop}%
\bibitem [{\citenamefont {Mak}\ \emph {et~al.}(2008)\citenamefont {Mak},
  \citenamefont {Sfeir}, \citenamefont {Wu}, \citenamefont {Lui}, \citenamefont
  {Misewich},\ and\ \citenamefont {Heinz}}]{HeinzAbsorbance2008}%
  \BibitemOpen
  \bibfield  {author} {\bibinfo {author} {\bibfnamefont {K.~F.}\ \bibnamefont
  {Mak}}, \bibinfo {author} {\bibfnamefont {M.~Y.}\ \bibnamefont {Sfeir}},
  \bibinfo {author} {\bibfnamefont {Y.}~\bibnamefont {Wu}}, \bibinfo {author}
  {\bibfnamefont {C.~H.}\ \bibnamefont {Lui}}, \bibinfo {author} {\bibfnamefont
  {J.~A.}\ \bibnamefont {Misewich}}, \ and\ \bibinfo {author} {\bibfnamefont
  {T.~F.}\ \bibnamefont {Heinz}},\ }\href {\doibase
  10.1103/PhysRevLett.101.196405} {\bibfield  {journal} {\bibinfo  {journal}
  {Phys. Rev. Lett.}\ }\textbf {\bibinfo {volume} {101}},\ \bibinfo {pages}
  {196405} (\bibinfo {year} {2008})}\BibitemShut {NoStop}%
\bibitem [{\citenamefont {Nair}\ \emph {et~al.}(2008)\citenamefont {Nair},
  \citenamefont {Blake}, \citenamefont {Grigorenko}, \citenamefont {Novoselov},
  \citenamefont {Booth}, \citenamefont {Stauber}, \citenamefont {Peres},\ and\
  \citenamefont {Geim}}]{NairGeim2008}%
  \BibitemOpen
  \bibfield  {author} {\bibinfo {author} {\bibfnamefont {R.~R.}\ \bibnamefont
  {Nair}}, \bibinfo {author} {\bibfnamefont {P.}~\bibnamefont {Blake}},
  \bibinfo {author} {\bibfnamefont {A.~N.}\ \bibnamefont {Grigorenko}},
  \bibinfo {author} {\bibfnamefont {K.~S.}\ \bibnamefont {Novoselov}}, \bibinfo
  {author} {\bibfnamefont {T.~J.}\ \bibnamefont {Booth}}, \bibinfo {author}
  {\bibfnamefont {T.}~\bibnamefont {Stauber}}, \bibinfo {author} {\bibfnamefont
  {N.~M.~R.}\ \bibnamefont {Peres}}, \ and\ \bibinfo {author} {\bibfnamefont
  {A.~K.}\ \bibnamefont {Geim}},\ }\href {\doibase 10.1126/science.1156965}
  {\bibfield  {journal} {\bibinfo  {journal} {Science}\ }\textbf {\bibinfo
  {volume} {320}},\ \bibinfo {pages} {1308} (\bibinfo {year}
  {2008})}\BibitemShut {NoStop}%
\bibitem [{\citenamefont {Ando}\ \emph {et~al.}(2002)\citenamefont {Ando},
  \citenamefont {Zheng},\ and\ \citenamefont
  {Suzuura}}]{AndoDynConductivity2002}%
  \BibitemOpen
  \bibfield  {author} {\bibinfo {author} {\bibfnamefont {T.}~\bibnamefont
  {Ando}}, \bibinfo {author} {\bibfnamefont {Y.}~\bibnamefont {Zheng}}, \ and\
  \bibinfo {author} {\bibfnamefont {H.}~\bibnamefont {Suzuura}},\ }\href
  {\doibase 10.1143/JPSJ.71.1318} {\bibfield  {journal} {\bibinfo  {journal}
  {J. Phys. Soc. Jpn.}\ }\textbf {\bibinfo {volume} {71}},\ \bibinfo {pages}
  {1318} (\bibinfo {year} {2002})}\BibitemShut {NoStop}%
\bibitem [{\citenamefont {Abergel}\ and\ \citenamefont
  {Fal'ko}(2007)}]{AbergelFalko2007}%
  \BibitemOpen
  \bibfield  {author} {\bibinfo {author} {\bibfnamefont {D.~S.~L.}\
  \bibnamefont {Abergel}}\ and\ \bibinfo {author} {\bibfnamefont {V.~I.}\
  \bibnamefont {Fal'ko}},\ }\href {\doibase 10.1103/PhysRevB.75.155430}
  {\bibfield  {journal} {\bibinfo  {journal} {Phys. Rev. B}\ }\textbf {\bibinfo
  {volume} {75}},\ \bibinfo {pages} {155430} (\bibinfo {year}
  {2007})}\BibitemShut {NoStop}%
\end{thebibliography}%


\begin{center}
      {\bf APPENDIX}
 \end{center}
\appendix
\section{Bloch Equations in RWA}
\label{Bloch Equation}
The optical response of semiconductors excited by coherent light sources are best described by the Optical Bloch Equation formalism. In presence of an optical field, the momentum couples to the field by Peierls minimal coupling: $\bm{p}\longrightarrow \bm{p}+e\bm{A}$ ($e>0$ is the magnitude of electronic charge). This implies that the total Hamiltonian in presence of the optical field is:
\begin{equation}
H_k=\hbar v_F \bm{\sigma}\cdot\bm{k}+ev_F \bm{\sigma}\cdot\bm{A}.
\end{equation}  
In the eigenbasis of the unperturbed Hamiltonian $H_o$, the complete Hamiltonian $H$ can be written as:

\begin{equation}
H_k=\sum_l \varepsilon_l \vert l \rangle \langle l\vert +\sum_{pq} e v_F \bm{A}\cdot\bm{\sigma}_{pq} \vert p\rangle \langle q\vert.
\end{equation}
where $\bm{\sigma}_{pq}= \langle p \vert \bm{\sigma} \vert q\rangle$. The sum over $l$, $p$ and $q$ $\in$ \{$c$,$v$\}.
The unperturbed Hamiltonian $H_o$:
\begin{equation}
H_o=\varepsilon_c \vert c \rangle \langle c\vert + \varepsilon_v \vert v \rangle \langle v\vert.
\end{equation}
The eigenbasis of the unperturbed Hamiltonian $H_o$ is: $\varepsilon_{c/v}=\pm \hbar v_F k$,
\begin{equation}
\vert c\rangle =\dfrac{1}{\sqrt{2}}\left(\begin{array}{c}
e^{-i\phi/2} \\ 
e^{i\phi/2} 
\end{array}\right) 
and \quad
\vert v\rangle =\dfrac{1}{\sqrt{2}}\left(\begin{array}{c}
e^{-i\phi/2} \\ 
-e^{i\phi/2} 
\end{array}\right)
\end{equation}
where $\phi=\text{tan}^{-1}\left(k_y/k_x \right)$.\\
The possible matrix elements $\bm{\sigma}_{pq}$ in the above basis can be evaluated:
\begin{equation}
\begin{array}{l}
\bm{\sigma}_{cc}=[\cos(\phi)\hat{x}+\sin(\phi)\hat{y}] \\ 
\bm{\sigma}_{vv}=-[\cos(\phi)\hat{x}+\sin(\phi)\hat{y}] \\ 
\bm{\sigma}_{cv}=[-i\sin(\phi)\hat{x}+i\cos(\phi)\hat{y}] \\ 
\bm{\sigma}_{vc}=[i\sin(\phi)\hat{x}-i\cos(\phi)\hat{y}]
\end{array} 
\end{equation}
Neglecting the intraband matrix elements (only interested in interband absorption), the perturbation Hamiltonian can be written as:
\begin{equation}
H^\prime=e v_F \bm{A}\cdot\bm{\sigma}_{cv} \vert c \rangle \langle v \vert +e v_F \bm{A}\cdot\bm{\sigma}_{vc} \vert v \rangle \langle c \vert.
\end{equation}
In the interaction representation (where the trivial time dependence from the unperturbed Hamiltonian is removed),
\begin{equation}
\begin{split}
\widetilde{H^\prime}&=\exp{\left( \dfrac{iH_o t}{\hbar}\right)} H^\prime \exp{\left( -\dfrac{i 
H_o t}{\hbar}\right)}\\
&=e v_F \bm{A}\cdot\bm{\sigma}_{cv} e^{i\omega_{cv}t}\vert c \rangle \langle v \vert 
+e v_F \bm{A}\cdot\bm{\sigma}_{vc} e^{-i\omega_{v}t} \vert v \rangle \langle c \vert
\end{split}
\end{equation}
where $\hbar\omega_{cv}=\varepsilon_c-\varepsilon_v$.
Assuming the vector potential of the optical field to be a pulse with a Gaussian envelope:
\begin{equation}
\bm{A}=\text{Re} \left[\bm{A_o} \exp{\left( -\dfrac{t^2}{2 w^2}\right)} \exp(-i\omega_o t)\right]
\label{GaussianVectorPotential}
\end{equation}
Thus in the RWA (Rotating Wave Approximation) i.e. keeping only near resonance terms and dropping fast oscillating terms,
\begin{equation}
\begin{split}
\widetilde{H^{\prime}_R}=&\dfrac{e v_F}{2} \bm{A_o}\cdot\bm{\sigma}_{cv}\exp{\left( -\dfrac{t^2}{2 w^2}\right)} e^{i\delta t}\vert c \rangle \langle v \vert \\
&+\dfrac{e v_F}{2} \bm{A_o}\cdot\bm{\sigma}_{vc} \exp{\left( -\dfrac{t^2}{2 w^2}\right)}e^{-i\delta t} \vert v \rangle \langle c \vert
\end{split}
\end{equation}
where $\delta=\omega_{cv}-\omega_o$.
From the Von-Neumann equation for the density matrix,
\begin{equation}
i\hbar \dfrac{\partial \widetilde{\rho}}{\partial t} = [\widetilde{H^\prime},\widetilde{\rho}],
\end{equation}
one can evaluate the time evolution of the density matrix components i.e the band occupation density (diagonal components of the density matrix) and the interband microscopic polarization (off-diagonal components of the density matrix).

\begin{equation}
\begin{split}
\dfrac{\partial \widetilde{\rho_{cc}}}{\partial t} &=-\dfrac{i e v_F}{2\hbar} e^{-t^2/2 w^2}\bm{A_o}\cdot \left[\bm{\sigma}_{cv} e^{i\delta t} \widetilde{\rho_{vc}}-\bm{\sigma}_{vc} e^{-i\delta t} \widetilde{\rho_{cv}}\right]\\
\dfrac{\partial \widetilde{\rho_{vv}}}{\partial t} &=\dfrac{i e v_F}{2\hbar} e^{-t^2/2 w^2}\bm{A_o}\cdot \left[\bm{\sigma}_{cv} e^{i\delta t} \widetilde{\rho_{vc}}-\bm{\sigma}_{vc} e^{-i\delta t} \widetilde{\rho_{cv}}\right]\\
\dfrac{\partial \widetilde{\rho_{cv}}}{\partial t} &=-\dfrac{i e v_F}{2\hbar} \bm{A_o}\cdot \bm{\sigma}_{cv} e^{-t^2/2 w^2} e^{i\delta t} \left( \widetilde{\rho_{vv}}-\widetilde{\rho_{cc}}\right)\\
\dfrac{\partial \widetilde{\rho_{vc}}}{\partial t}&=\,\dfrac{i e v_F}{2\hbar} \bm{A_o}\cdot \bm{\sigma}_{vc} e^{-t^2/2 w^2} e^{-i\delta t} \left( \widetilde{\rho_{vv}}-\widetilde{\rho_{cc}}\right)
\end{split}
\end{equation}

\subsection{Undoped Graphene Absorption}
\label{Undoped Graphene Absorption}

The velocity operator for graphene in the Dirac dispersion approximation is $\vec{v}=v_F \vec{\sigma}$. \\

This set of equations explain the absorption of undoped graphene. In case of monochromatic vector potential: $\vec{A}=\text{Re}[\vec{A}_o e^{-i\omega_o t}]$, the optical field vector potential is $\vec{A}=\vec{A}_o [e^{-i\omega_o t}+e^{i \omega_o t}]/2$. Thus $\vec{A}(\omega_o)=\vec{A}_o /2$.
The susceptibility is defined by: 
\begin{equation}
\begin{split}
\chi (\omega)&=\dfrac{j(\omega)}{\varepsilon_o \omega^2 A(\omega)}\\
&=\dfrac{e^2v_F^2}{\varepsilon_o \omega^2 \hbar L^2} \sum_{\vec{k}} \sin^2(\phi) \left(\dfrac{f_v-f_c}{\omega_{cv}-\omega-i\eta} \right)
\end{split}
\end{equation}
Assuming valence band to be full and conduction band to be empty, $f_v=1$ and $f_c=0$.
\begin{equation}
\text{Im}[\chi(\omega)]=\dfrac{\pi e^2v_F^2}{\varepsilon_o \omega^2 \hbar L^2} \sum_{\vec{k}} \sin^2(\phi) \delta(\omega_{cv}-\omega)
\end{equation}
 The absorption relates to the imaginary part of the susceptibility.
\begin{equation}
\begin{split}
\alpha(\omega)&=\dfrac{\omega}{c} \text{Im}[\chi(\omega)]\\
&=\dfrac{ e^2}{ 4 \varepsilon_o \hbar c}=0.0231
\end{split}
\end{equation}

This is consistent with the measured and calculated interband absorption from far infrared to ultraviolet spectrum range \cite{HeinzAbsorbance2008,NairGeim2008,AndoDynConductivity2002,AbergelFalko2007}.


\subsection{Generation Rate under weak pump excitation }
\label{Generation Rate}

Considering the case of weak pump excitation $\left( \dfrac{e v_F A_o}{2 \hbar}\ll 1\right)$ for the pulse with Gaussian envelope such that to lowest order: 
\begin{equation}
\begin{split}
\dfrac{\partial \widetilde{\rho_{cv}}}{\partial t} &=-\dfrac{i e v_F}{2\hbar} \vec{A_o}\cdot \vec{\sigma}_{cv} \exp{\left( -\dfrac{t^2}{2 w^2}\right)}e^{i\delta t} \left( \widetilde{\rho_{vv}}-\widetilde{\rho_{cc}}\right)\\
&\simeq -\dfrac{i e v_F}{2\hbar} \vec{A_o}\cdot \vec{\sigma}_{cv} \exp{\left( -\dfrac{t^2}{2 w^2}\right)}e^{i\delta t} \left( {\rho_{vv}}-{\rho_{cc}}\right)\\
 \widetilde{\rho_{cv}}&\simeq -\dfrac{i e v_F}{2\hbar} \vec{A_o}\cdot \vec{\sigma}_{cv}  \left( {\rho_{vv}}-{\rho_{cc}}\right) \int_{-\infty}^{t} dt' \exp{\left( -\dfrac{t'^2}{2 w^2}\right)}e^{i\delta t'}\\
&\simeq -\sqrt{\dfrac{\pi}{2}}w\dfrac{i e v_F}{2\hbar} \vec{A_o}\cdot \vec{\sigma}_{cv}  \left( {\rho_{vv}}-{\rho_{cc}}\right)\exp{\left( -\dfrac{\delta^2 w^2}{2}\right)}\\
&\qquad \times \left[ 1+\text{Erf}\left( \dfrac{t}{\sqrt{2}w}-\dfrac{i\delta w}{\sqrt{2}}\right)\right]
\end{split}
\end{equation}
Substituting the above expression in the equation for the diagonal component of the density matrix with the vector potential polarization $\vec{A_o}=A_o \hat{\lambda}$, 
\begin{equation}
\begin{split}
\dfrac{\partial f_c}{\partial t}&=\dfrac{\partial \widetilde{\rho_{cc}}}{\partial t} \\
&=\sqrt{\dfrac{\pi}{2}}\dfrac{e^2 v_F^2 w}{2 \hbar^2}A_o^2 \vert \sigma_{cv}^{\lambda}\vert^2 \exp{\left(-\dfrac{t^2}{2 w^2}\right)} \exp{\left(-\dfrac{w^2 \delta^2}{2}\right)}\\
&\times (f_v-f_c) \text{Real}\left[ e^{-i\delta t}\left( 1+\text{Erf}\left( \dfrac{t}{\sqrt{2}w}-\dfrac{i\delta w}{\sqrt{2}}\right)\right)\right]
\end{split}
\end{equation}
In using the generation rate for further calculations, we assume that $f_c=0$ and $f_v=1$.


\section{Boltzmann Equation Solution}
\label{Solution to BTE}
The BTE in Eq.~(\ref{BTEGeneration}) can be solved by Fourier transforms of the distribution function defined by:
\begin{eqnarray}
 \nonumber &&
g( \bm{r} )=\dfrac{1}{2 \pi} \iint e^{-i\bm{k}\cdot\bm{r}} f( \bm{k} )\,d\bm{k} 
\\  &&
f( \bm{k} )=\dfrac{1}{2 \pi} \iint e^{ i\bm{k}\cdot\bm{r}}g( \bm{r} )\,d\bm{r}
\label{FT}
\end{eqnarray}
Upon Fourier transforming the BTE in Eq.~(\ref{BTEGeneration}):
\begin{equation}
\begin{split}
\dfrac{\partial g}{\partial t}+\dfrac{ieEx}{\hbar}g
&=\dfrac{\partial g_g}{\partial t}-\dfrac{g-g_t}{\tau_T}-\dfrac{g-g_c}{\tau_C}\\
\Rightarrow \dfrac{\partial }{\partial t}[g e^{t/\tau}e^{ieExt/\hbar}]&=e^{t/\tau}e^{ieExt/\hbar}\left(\dfrac{\partial g_g}{\partial t}+\dfrac{g_t}{\tau_T}+\dfrac{g_c}{\tau_C}\right)
\end{split}
\end{equation}
where $\tau^{-1}=\tau_T^{-1}+\tau_C^{-1}$. Integrating over time from an initial time $t_i$ (well before the applied pump pulse) gives the solution in Fourier space:
\begin{equation}
\begin{split}
g=&g_{i} e^{-(t-t_i)/ \tau}e^{-ieEx(t-t_i)/ \hbar}+e^{-t/ \tau} e^{-ieExt/\hbar} \\
&\times \int_{t_i}^{t}dt^\prime e^{t^\prime/ \tau} e^{ieExt^\prime/\hbar} \left(\dfrac{\partial g_g}{\partial t^\prime}+\dfrac{g_t}{\tau_T}+\dfrac{g_c}{\tau_C}\right)\\
\end{split}
\label{Solution_BTEGeneration}
\end{equation}
where $g_i$ in the Fourier transform of the steady state distribution before the pump is applied, which means the steady-state distribution of the carriers in presence of DC electric field due to thermal excited carriers/initial doping. $g_g$/$g_t$/$g_c$ are the Fourier transforms of $f_g$/$f_t$/$f_c$. Upon taking the inverse Fourier transform, one can get the time dependent distribution function. However, the initial steady-state distribution function before photoexcitation needs to be determined.\\

The carriers prior to the optical pump pulse are in a steady state which can be calculated using the Boltzmann equation. The steady state Boltzmann equation for the initial distribution prior to the pump pulse is one similar to Eq.~(\ref{BTEGeneration}) except that the time derivative term is absent since steady state and the generation rate term is absent since the system prior to photoexcitation is being considered:
\begin{equation}
\dfrac{eE}{\hbar}\dfrac{\partial f_i}{\partial k_x}=-\dfrac{f_i-f_{it}}{\tau_T}-\dfrac{f_i-f_{ic}}{\tau_C}
\label{InitialBTE}
\end{equation}
where the subscript $i$ in the distribution functions corresponds to `initial'. The collision terms are the same as stated in Eq.~(\ref{Thermalization}) and Eq.~(\ref{Cooling}).   
\begin{equation}
\begin{split}
f_{it}&=[1+\exp\{(\hbar v_F \vert \bm{k}-\bm{k_d}^i\vert-\mu^{*i})/(k_b T_{el}^i)\}]^{-1}\\
f_{ic}&=[1+\exp\{(\hbar v_F \vert \bm{k}\vert-\mu^{i})/(k_b T)\}]^{-1}.
\label{Initial_Collision_Integral}
\end{split}
\end{equation}
The moments of the BTE (Eq.~(\ref{InitialBTE})) for the initial distribution with respect to wavevector $k_x$ and energy $\varepsilon_{\bm{k}}$ is given by:
\begin{equation}
\begin{split}
k_d^i&=\dfrac{\langle k_x \rangle}{n}=\dfrac{eE\tau_C}{\hbar}\\
\langle \varepsilon_k \rangle^i&=\langle \varepsilon_c\rangle^i + eE \tau_C \langle v_x \rangle_n^i\\
\end{split}
\end{equation}
\begin{equation}
\begin{split}
\text{where}\; \langle v_x \rangle_n^i&=\dfrac{1}{(2 \pi)^2} \iint v_x(\bm{k}) f_i( \bm{k} )\,d\bm{k} \\
 \langle \varepsilon_k \rangle^i&=\dfrac{1}{(2 \pi)^2} \iint \varepsilon_{\bm{k}} f_i( \bm{k} )\,d\bm{k} \\  \langle \varepsilon_c \rangle^i&=\dfrac{1}{(2 \pi)^2} \iint \varepsilon_{\bm{k}} f_c( \bm{k} )\,d\bm{k} .
\end{split}
\end{equation}
The solution to the BTE (Eq.~(\ref{InitialBTE})) in Fourier space is:
\begin{equation}
g_i=\dfrac{\left[\dfrac{\tau}{\tau_T}g_{it} +\dfrac{\tau}{\tau_C}g_{ic}\right]}{1+ieEx\tau/\hbar}
\end{equation}
The energy moment relation suggests that the initial steady state distribution needs to be solved self-consistently since the energy is related to the average velocity of the carriers. Thus starting with a guess of average energy $\langle\varepsilon_{\bm{k}}\rangle^i$ (initial carrier number density $n_i$ is fixed ), $\mu^{*i}$ and $T_{el}^i$ is solved for and then the initial steady state distribution is evaluated. This initial distribution is used to find the average energy and this procedure is repeated until self-consistency is achieved.\\

Thus solution to the BTE (\ref{Solution_BTEGeneration}) requires the time-dependent parameters $k_d$, $\mu$, $\mu^*$, $T_{el}$ in the collision integral. These can be determined from the moments of the BTE as described below.

\section{Moments of distribution function }
\label{Moments}

\subsection{Moment equation for Carrier Density}
Since the thermalization and cooling collision integrals conserves number density, one can integrate the BTE (Eq.~(\ref{BTEGeneration})) over all $\bm{k}$ states:
\begin{equation}
\dfrac{\partial n}{\partial t}=\dfrac{\partial n_g}{\partial t}=\iint \dfrac{d\bm{k}}{(2\pi)^2} \dfrac{\partial f_g}{\partial t}
\end{equation}
The equation can be integrated in time to get:
\begin{equation}
n(t)=n(t_i)+\int_{t_i}^{t}dt^\prime \dfrac{\partial n_g}{\partial t^\prime}
\end{equation}

\subsection{Moment equation for Wavevector}

To determine the time evolution of the drift wave-vector, one can multiply the BTE (Eq.~(\ref{BTEGeneration})) by $k_x$ and then integrate over all $\bm{k}$ states:
\begin{equation}
\dfrac{\partial k_d^n}{\partial t}-\dfrac{e E n}{\hbar}=-\dfrac{k_d^n}{\tau_C}
\end{equation}
where $k_d^n=\langle k_x \rangle$ is the average of wavevector over the distribution function (Average of $n$ carriers). Integration of the above equation gives a time-dependent drift wave-vector:
\begin{equation}
k_d^n(t)=k_d^n(t_i) e^{-(t-t_i)/\tau_C}+\dfrac{e E}{\hbar} e^{-t/\tau_C} \int_{t_i}^{t} dt^\prime e^{t^\prime/\tau_C} n(t^\prime)
\end{equation}
Thus the drift wave-vector of the distribution $k_d(t)=k_d^n(t)/n(t)$.\\

\begin{figure}[htbp]
\centering
\includegraphics[scale=0.35]{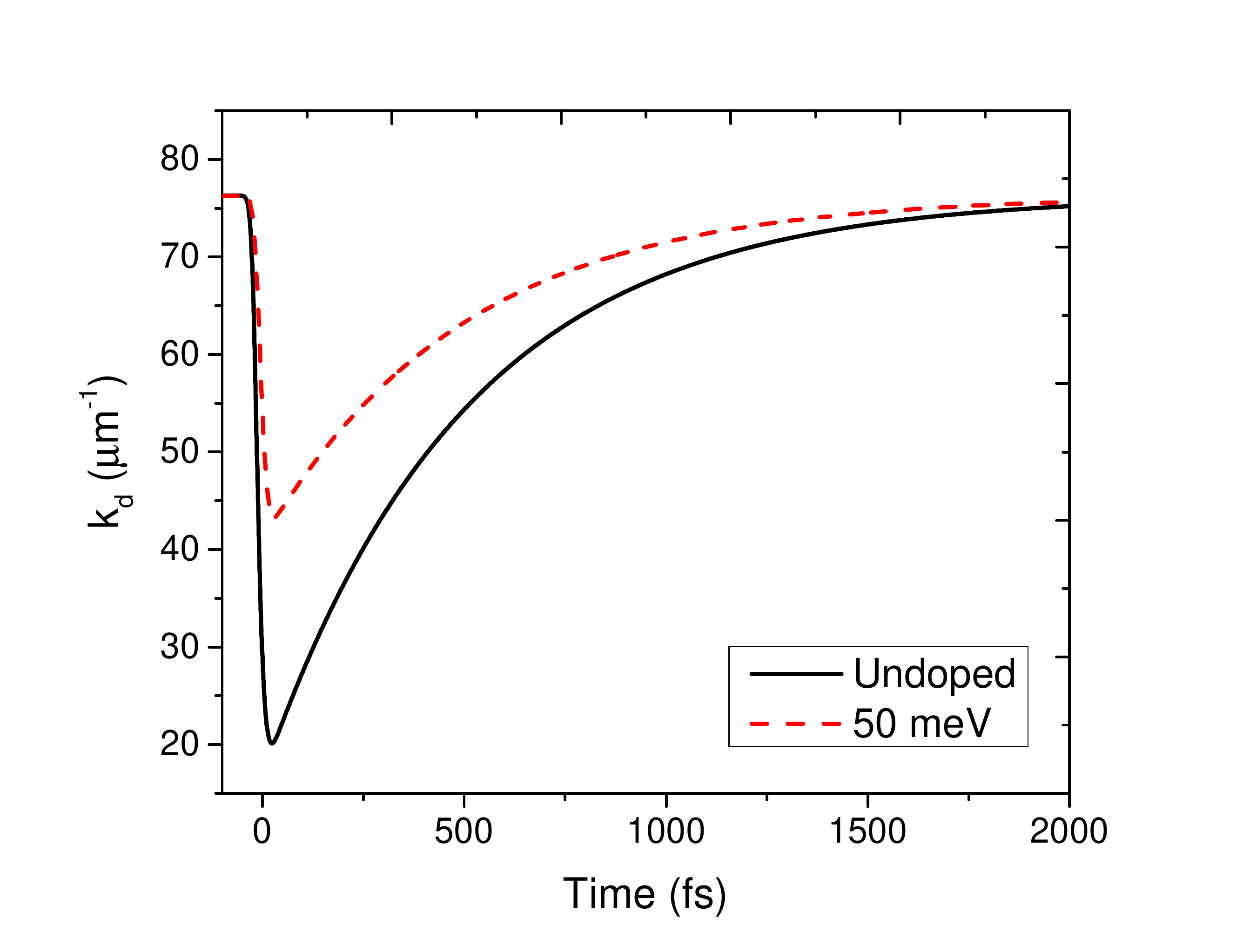}
\caption{\label{drift_wavevector}(color online) Drift wavevector of a single carrier as a function of time for undoped and for n-doped graphene with room temperature chemical potential of 50 meV.}
\end{figure}

The time variation of the drift wavevector $k_d$ per carrier of the drifted Fermi-Dirac distribution is shown in Fig.~\ref{drift_wavevector}. The thermalization collision integral attributed to scattering between carriers relaxes the distribution to a drifted Fermi-Dirac distribution with an effective temperature and quasi-Fermi level with the drift wavevector $k_d$. This drift wavevector as seen decreases during the pulse width duration as the number of the carriers increases in that time duration but the carriers are generated with no wavevector by the optical pump. After the pump pulse is over the distribution starts to drift under the effect of DC electric field.

\subsection{Moment equation for Energy Density}
To determine the time evolution of energy, one can multiply the BTE (Eq.~(\ref{BTEGeneration})) by $\varepsilon_k$ and then integrate over all $\bm{k}$ states:
\begin{equation}
\dfrac{\partial \langle\varepsilon_{\bm{k}}\rangle}{\partial t}-e E \langle v_x \rangle_n=\dfrac{\partial \langle \varepsilon_g\rangle}{\partial t}-\dfrac{\langle\varepsilon_{\bm{k}}\rangle-\langle\varepsilon_{c}\rangle}{\tau_C}
\end{equation}
where $\langle v_x \rangle_n$ is the average of velocity over the distribution function $f$ (Average of $n$ carriers), $\langle \varepsilon_{\bm{k}} \rangle$ is the average of energy over the distribution function $f$, $\langle \varepsilon_{c} \rangle$ is the average of energy over the cooling distribution function $f_c$, and $\partial_t \langle \varepsilon_g\rangle$ is the average of energy over the generation rate $\partial_t f_g$. Integration of the above equation gives a time-dependent energy:
\begin{equation}
\begin{split}
\langle\varepsilon_{\bm{k}}\rangle(t)=&\langle\varepsilon_{\bm{k}}\rangle(t_i) e^{-(t-t_i)/\tau_C}\\
&+ e^{-t/\tau_C} \int_{t_i}^{t} dt^\prime e^{t^\prime/\tau_C} \left[\dfrac{\partial \langle \varepsilon_g \rangle}{\partial t^\prime} + eE\langle v_x \rangle_n +\dfrac{\langle \varepsilon_c \rangle}{\tau_C}\right]
\end{split}
\end{equation}

\begin{figure}[htbp]
\centering
\includegraphics[scale=0.35]{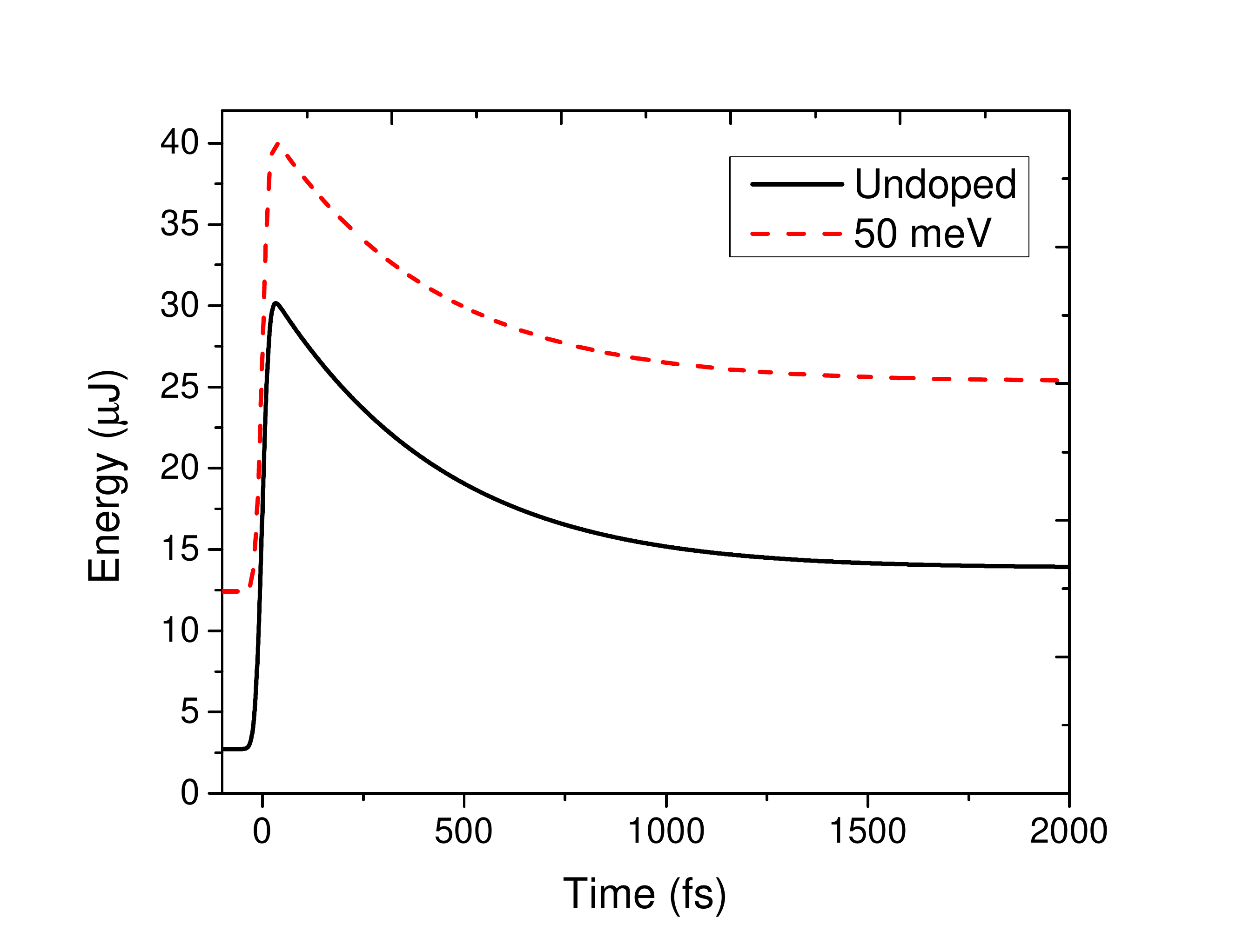}
\caption{\label{Energy}(color online) Energy density of carriers as a function of time for undoped and for n-doped graphene with room temperature chemical potential of 50 meV.}
\end{figure}

The average energy of the carriers increase during generation of carriers since the pump provides energy to photoexcited carriers and thermalization mainly corresponds to carrier-carrier interaction which conserves energy as seen in Fig.~\ref{Energy}. After the thermalization time scale the carriers begin to cool down by giving their energy to the lattice.\\

\begin{figure}[htbp]
\centering
\includegraphics[scale=0.35]{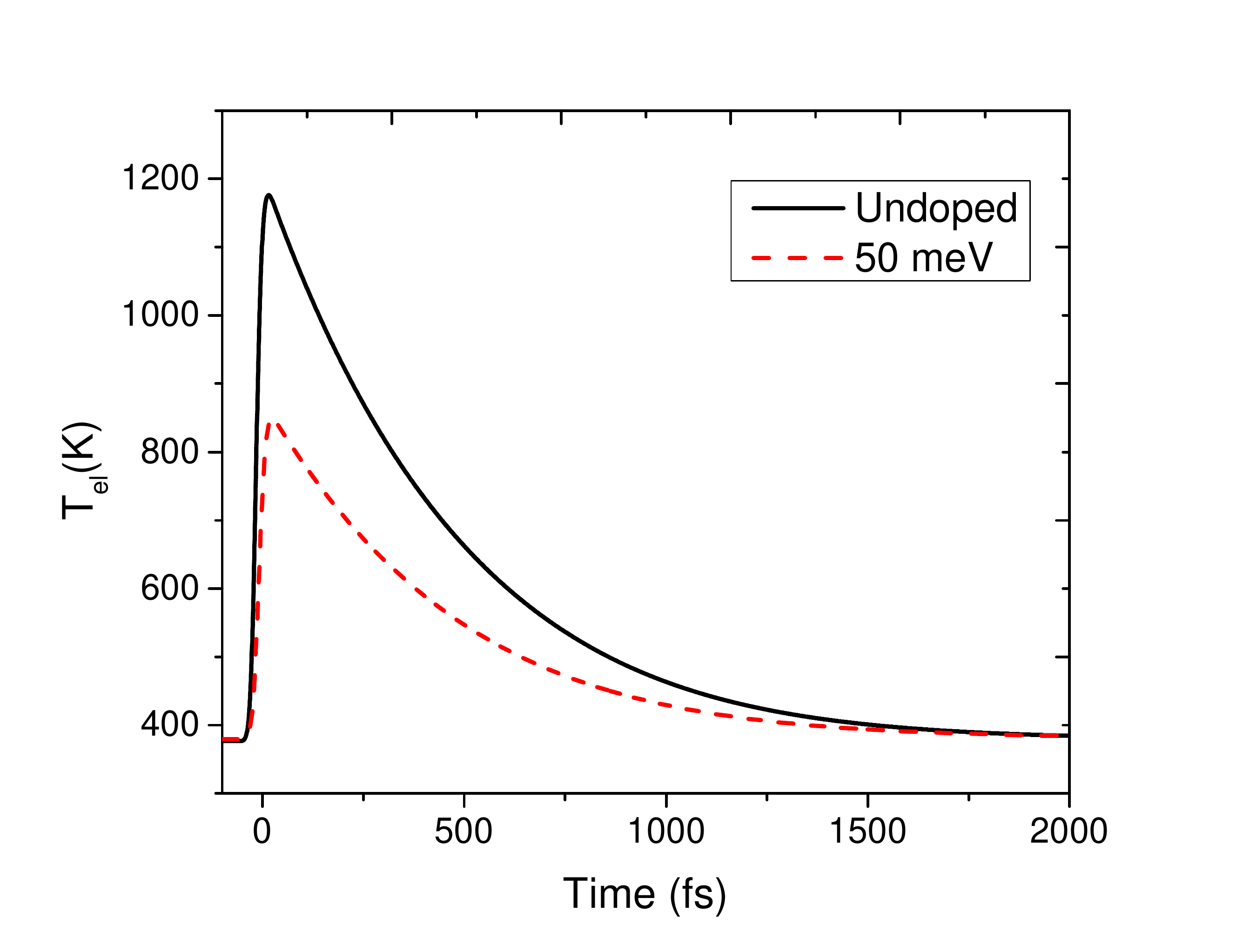}
\caption{\label{electron_temperature}(color online) Electron temperature of thermalized distribution as a function of time for undoped and for n-doped graphene with room temperature chemical potential of 50 meV. Note that the electron temperature is higher than the lattice temperature (T=300 K) because of the heating of the electrons due to the DC field.}
\end{figure}
\begin{figure}[htbp]
\centering
\includegraphics[scale=0.35]{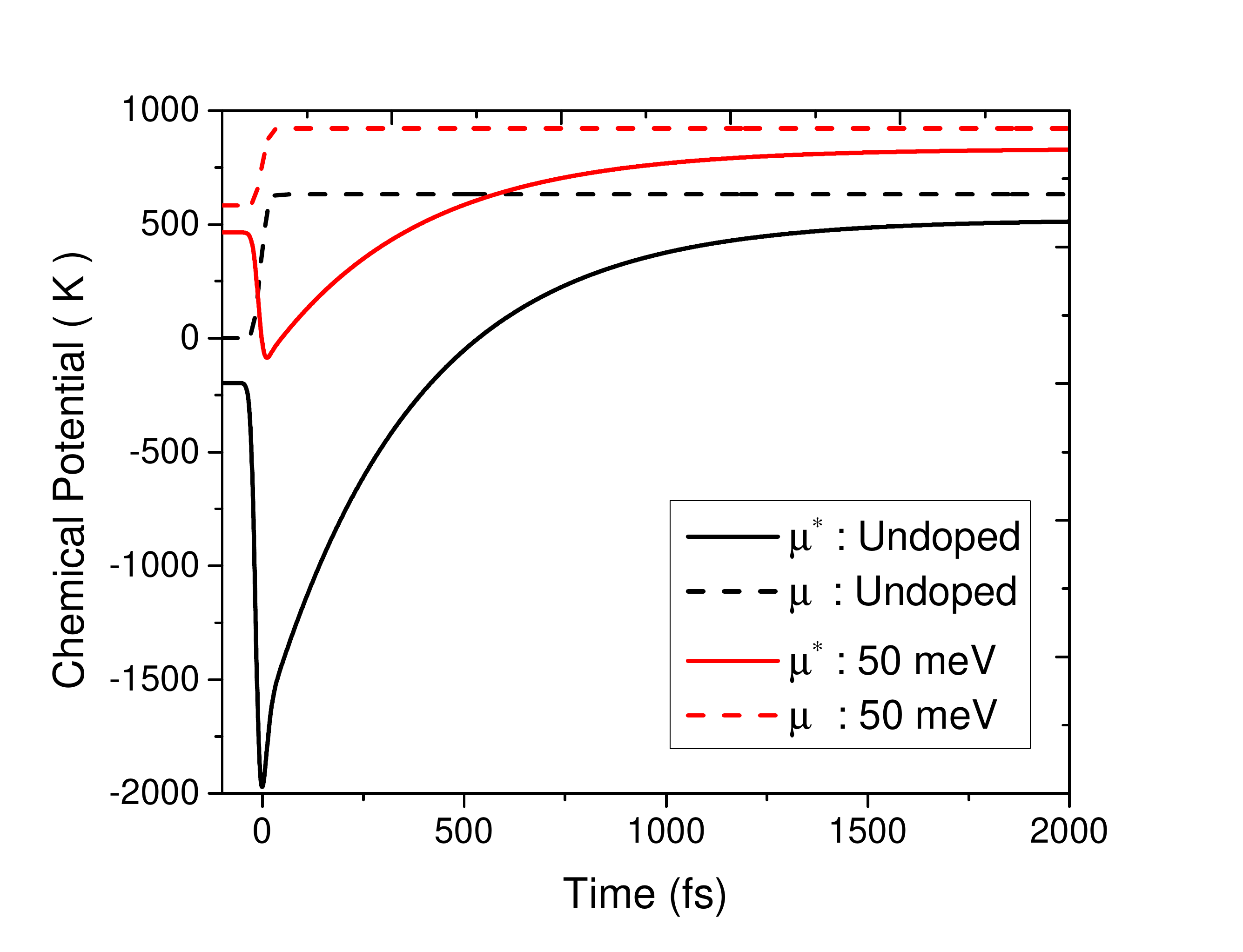}
\caption{\label{chemical_potential}(color online) Chemical potential for the thermalization and cooling distributions as a function of time for undoped and for n-doped graphene with room temperature chemical potential of 50 meV.}
\end{figure}

The time-dependent wave-vector ($k_d$), number density ($n$), and energy ($\langle\varepsilon_{\bm k}\rangle$) determine the parameters ($\mu^*$,$T_{el}$ and $\mu$). The time evolution of the effective electron temperature follows the average energy as seen in Fig.~\ref{electron_temperature}. This trend is seen in the effective carrier temperature in the thermalization collision integral. The electron temperature increases as energy is pumped though photoexcitation. The electron temperature increases as the photoexcited carrier density increases and eventually cools down to the initial steady-state electron temperature since the presence of the DC electric field increases the steady-state temperature of the thermally excited/doped carriers prior to the optical pump pulse.\\

The chemical potential for the cooling collision integral follows the number density as seen in Fig.~\ref{chemical_potential}. However the quasi-Fermi level for the thermalization collision integral $\mu^*$ initially decreases. This happens because the effective electron temperature increases due to the energy pumped into the system by the pump pulse but the conservation of the instantaneous number density requires the quasi-Fermi level $\mu^*$ to decrease. After the duration of the pump pulse, the quasi-Fermi level $\mu^*$ increases as the carriers cool down.\\

\end{document}